# Compositional distributions and evolutionary processes for the near-Earth object population: Results from the MIT-Hawaii Near-Earth Object Spectroscopic Survey (MITHNEOS)

R.P. Binzel[a,1,*], F.E. DeMeo[a], E.V. Turtelboom[a], S.J. Bus[b], A. Tokunaga[b], T.H. Burbine[c], C. Lantz[d], D. Polishook[e], B. Carry[f], A. Morbidelli[f], M. Birlan[g], P. Vernazza[h], B.J. Burt[i], N. Moskovitz[i], S.M. Slivan[j], C.A. Thomas[k], A.S. Rivkin[l], M.D. Hicks[m], T. Dunn[n], V. Reddy[o], J.A. Sanchez[p], M. Granvik[q], T. Kohout[r]

[a] *Department of Earth, Atmospheric, and Planetary Sciences, Massachusetts Institute of Technology, Cambridge, MA, United States* [b] *University of Hawaii, United States* [c] *Mount Holyoke College, United States*
[d] *Institut d'Astrophysique Spatiale, CNRS/Université Paris Saclay, France* [e] *Weizmann Institute, Israel*
[f] *Université Côte d'Azur, Observatoire de la Côte d'Azur, CNRS, Lagrange, France*
[g] *IMCCE, Paris Observatory and Astronomical Institute of Romanian Academy, Bucharest, Romania* [h] *LAM–CNRS/AMU, Marseille, France* [i] *Lowell Observatory, United States* [j] *Wellesley College, United States* [k] *Northern Arizona University, United States*
[l] *Johns Hopkins University Applied Physics Laboratory, United States* [m] *Jet Propulsion Laboratory, United States* [n] *Colby College, United States*
[o] *University of Arizona, United States* [p] *Planetary Science Institute, United States*
[q] *University of Helsinki and Luleå University of Technology, Kiruna, Sweden* [r] *University of Helsinki, Finland*

* Corresponding author.
[1] Chercheur Associé, IMCCE-Observatoire de Paris, France.

ABSTRACT

Advancing technology in near-infrared instrumentation and dedicated planetary telescope facilities have enabled nearly two decades of reconnoitering the spectral properties for near-Earth objects (NEOs). We report measured spectral properties for more than 1000 NEOs, representing >5% of the currently discovered population. Thermal flux detected below 2.5 μm allows us to make albedo estimates for nearly 50 objects, including two comets. Additional spectral data are reported for more than 350 Mars-crossing asteroids. Most of these measurements were achieved through a collaboration between researchers at the Massachusetts Institute of Technology and the University of Hawaii, with full cooperation of the NASA Infrared Telescope Facility (IRTF) on Mauna Kea. We call this project the MIT-Hawaii Near-Earth Object Spectroscopic Survey (MITHNEOS; myth-neos). While MITHNEOS has continuously released all spectral data for immediate use by the scientific community, our objectives for this paper are to: (1) detail the methods and limits of the survey data, (2) formally present a compilation of results including their taxonomic classification within a single internally consistent framework, (3) perform a preliminary analysis on the overall population characteristics with a concentration toward deducing key physical processes and identifying their source region for escaping the main belt. Augmenting our newly published measurements are the previously published results from the broad NEO community, including many results graciously shared by colleagues prior to formal publication. With this collective data set, we find the near-Earth population matches the diversity of the main-belt, with all main-belt taxonomic classes represented in our sample. Potentially hazardous asteroids (PHAs) as well as the subset of mission accessible asteroids (ΔV≤7 km/s) both appear to be a representative mix of the overall NEO population, consistent with strong dynamical mixing for the population that interacts most closely with Earth. Mars crossers, however, are less diverse and appear to more closely match the inner belt population from where they have more recently diffused. The fractional distributions of major taxonomic classes (60% S, 20% C, 20% other) appear remarkably constant over two orders of magnitude in size (10 km to 100 m), which is eight orders of magnitude in mass, though we note unaccounted bias effects enter into our statistics below about 500 m. Given the range of surface ages, including possible refreshment by planetary encounters, we are able to identify a very specific space weathering vector tracing the transition from Q- to Sq- to S-types that follows the natural dispersion for asteroid spectra mapped into principal component space. We also are able to interpret a shock darkening vector that may account for some objects having featureless spectra. Space weathering effects for C-types are complex; these results are described separately by Lantz, Binzel, DeMeo. (2018, Icarus 302, 10–17). Independent correlation of dynamical models with taxonomic classes map the escape zones for NEOs to main-belt regions consistent with well established heliocentric compositional gradients. We push beyond taxonomy to interpret our visible plus near-infrared spectra in terms of the olivine and pyroxene mineralogy consistent with the H, L, and LL classes of ordinary chondrites meteorites. Correlating meteorite interpretations with dynamical escape region models shows a preference for LL chondrites to arrive from the ν6 resonance and H chondrites to have a preferential signature from the mid-belt region (3:1 resonance). L chondrites show some preference toward the outer belt, but not at a significant level. We define a Space Weathering Parameter as a continuous variable and find evidence for step-wise changes in space weathering properties across different planet crossing zones in the inner solar system. Overall we hypothesize the relative roles of planetary encounters, YORP spin-up, and thermal cycling across the inner solar system.

## 1. Introduction

Near-Earth objects (NEOs) are our planet's closest neighbors in space, simultaneously providing tantalizing opportunities for advancing scientific knowledge of our solar system, beckoning as exploration and resource utilization destinations, while at the same time threatening civilization by their potential for catastrophic impacts. Telescopic characterization is a key component for advancing all aspects of NEO understanding for both science and planetary protection. In this work we concentrate on the spectral (color) characterization of NEOs as a step toward interpreting their compositions and origins, exploration potential, and hazard assessment. We present the results of a long-term collaborative partnership that we call the MIT-Hawaii Near-Earth Object Spectroscopic Survey (abbreviated MITHNEOS; *myth-neos*). Researchers at the Massachusetts Institute of Technology (MIT) served as the organizational leaders with the University of Hawaii supporting the staffing and operations for the NASA Infrared Telescope Facility on Mauna Kea. While immediate release of the spectral measurements for open use by the scientific community has been a central tenet of the survey's operation since inception, formal systematic classification and interpretive analysis of the data set *en masse* has been forestalled pending accumulation of a sizeable sample for statistically robust investigation.

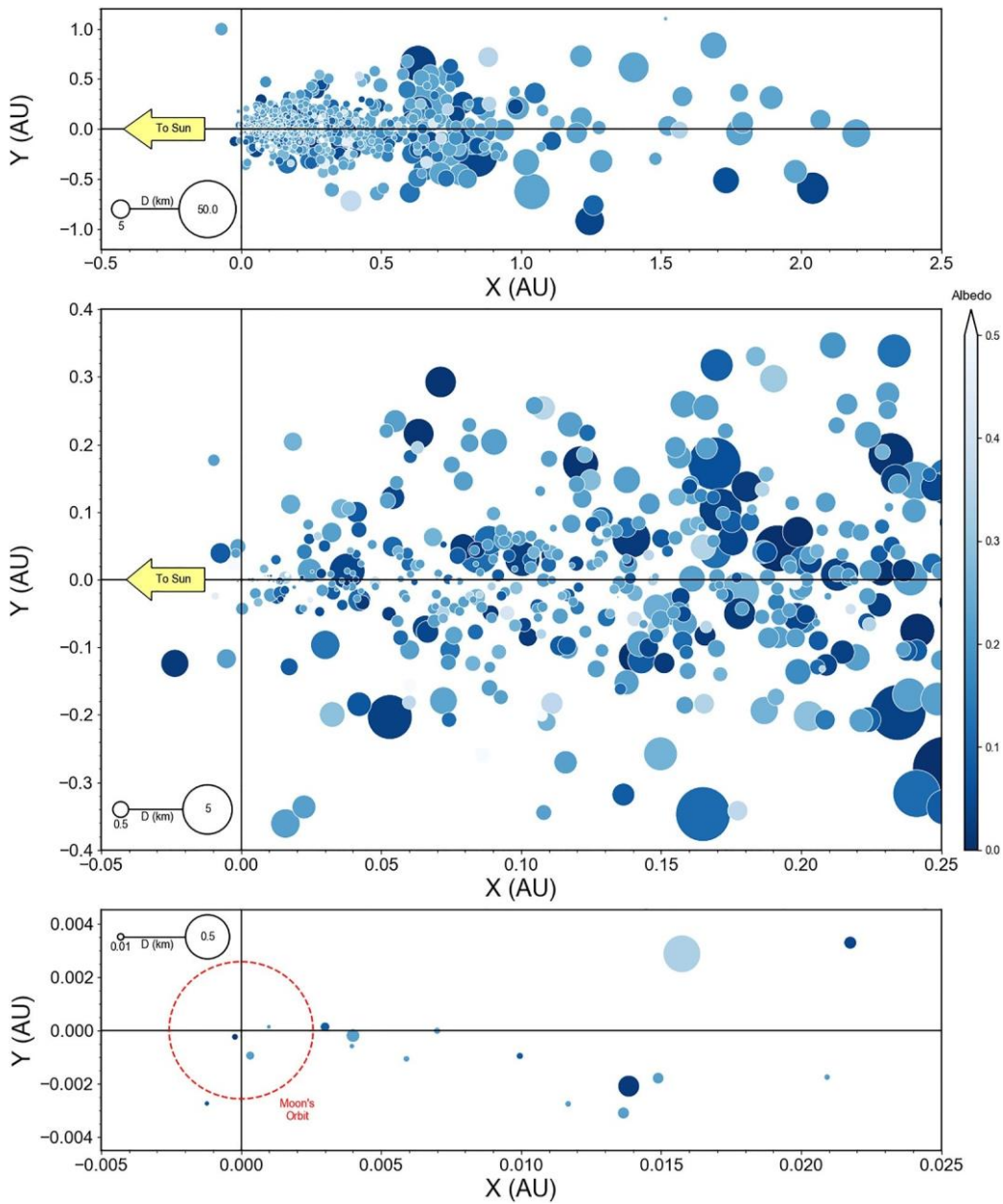

**Fig. 1.** Inner solar system position of all MITHNEOS targets at the time of their spectral measurement (geocentric X, Y coordinates in astronomical units [AU]). Each panel shows a factor of 10 closer view near Earth's location at the axes origin. Circle size depicts the estimated diameter of each object according to the scale at left for each panel. Shading scale indicates measured or inferred albedo according to taxonomic class.

We begin with a discussion of the definitions we use for describing the NEO population. In Section 2 we describe the survey methods for our observations and outline our tabulation of results. We also describe how complementary data sets from the literature have been incorporated as an augmentation to the tabulation, noting our indebtedness to members of the planetary astronomy community who very generously made available in digital form published (and in some cases unpublished) measurements supporting the completeness of this analysis. Our tabulation in Section 3 and Appendix II endeavors to fully attribute all of these sources. Classification of NEO spectral measurements is detailed in Section 3, where we explore the limits of our survey arising from observational uncertainties and identifiable systematic effects. We also explore trends arising from the exogenic effects of space weathering and impact-induced shock darkening. For the present work, our analysis beginning in Section 4 illuminates the diversity for both the NEO population and its subclasses within the context of basic asteroid taxonomic classes. In Section 5 we push beyond spectral curves and taxonomic classes to the next steps of compositional interpretation by forging links with the most predominantly falling meteorite classes (the ordinary chondrites). We seek to advance our understanding of the NEO population by correlating their compositional properties with their escape regions for delivery into the near-Earth space in Section 6. We push our meteorite mineralogical analysis and dynamical correlations to their limits to investigate the main belt escape regions for H, L, and LL classes of ordinary chondrites meteorites in Section 7. We use a quantitative estimate for space weathering to evaluate multiple processes that may resurface NEOs in the planet crossing region of the inner solar system in Section 8, and form a brief summary of conclusions and future work in Section 9.

*1.1. NEO population definitions*

Given the large size of our data set and the different ways it may be examined, we begin by defining our selection criteria for inclusion. For consistency, we follow the sub-division of the population in terms of orbital properties as suggested by Shoemaker et al. (1979). To be classified as a near-Earth object, the orbit perihelion distance (q) must be less than 1.3 astronomical units (AU). Amor objects (abbreviated AMO) are those whose perihelia are greater than 1.017 AU but less than 1.3 AU. Apollo objects (abbreviated APO) have semi-major axes (a) greater than 1 AU with $q \leq 1.017$ AU. Aten objects (abbreviated ATE) intersect with Earth's orbit from 'the inside out' by having a 〈 1 AU and aphelia greater than 0.983 AU. Atira objects (abbreviated ATI; but not sampled in our survey) have orbits entirely interior to Earth's. Our survey also includes Mars crossing (MC) objects, where we follow the methodology of Binzel et al. (2015) in setting the MC criteria as: a > 1.3 AU and $1.3 \leq q < 1.665$ AU, where the larger value is roughly the aphelion distance of Mars. We use current osculating orbital elements for our definitions, noting that objects near the numerical boundaries will periodically phase in and out of their categories over long orbital evolution timescales.

We also note here our preference to refer to this total population as near-Earth objects (NEOs) rather than near-Earth asteroids (NEAs) so as to recognize that both asteroid and comet sources can contribute to the population. We address this further in Section 4.

**2. MITHNEOS observations with the NASA IRTF and MIT Magellan telescope**

We present thumbnail versions of our reported spectral results in Appendix I and give their full tabulation in Appendix II. New spectral data for more than 750 objects are contributed by this study, where Fig. 1 depicts their geocentric viewing geometry on their date of measurement. The majority of these new spectral data were obtained using the instrument "SpeX" (Rayner et al., 2003) on the 3-meter NASA Infrared Telescope Facility (IRTF) located on Mauna Kea, Hawaii. As has been long documented (e.g. Gaffey et al., 1993), the capability to reveal and measure spectral characteristics of absorption bands present near 1 and 2-microns provides a highly diagnostic tool for distinguishing taxonomic classes and for enabling follow-on steps of compositional interpretation. We operated the SpeX instrument in its low resolution mode allowing the simultaneous collection of photons spanning the wavelength range of 0.8–2.5 μm. About fifty of the new measurements covering the same near-infrared wavelength range were obtained with the 6.5 m Baade Telescope at the Magellan Observatory in Las Campanas, Chile, using the FIRE (Folded-port InfraRed Echellette) instrument described by Simcoe et al. (2013). Magellan observing and reductions are detailed by Binzel et al. (2015), noting that we employed essentially identical observation and reduction techniques for the two telescopic systems.

Our desired telescopic data product for each target object is a reflectance spectrum, defined by calculating the ratio between the reflected light emanating from the asteroid (numerator) and the solar flux incident on the asteroid (denominator). Most ideally, this data product would be achieved by simultaneous measurement of the asteroid's reflected light *and the Sun* through the same optics and with the same detectors. In this ideal way through the same instrumentation, the unique characteristics of the optical transmission functions and detector response functions would precisely cancel in the mathematical quotient yielding each reflectance spectrum. Alas, the mutually exclusive reality of nighttime telescopic observing and the huge dynamic range that would be demanded to record the direct solar flux preclude this ideal. Thus we followed the practice of nighttime measurement of stars known to be very close spectral analogs to the Sun itself. Our primary solar analog standard stars were 16 Cyg B (SAO 31899) and Hyades 64 (SAO 93936), and when not available in the sky, additional solar analog stars were utilized that were verified as excellent matches in their flux versus wavelength spectral curves (to within ~1%) when ratioed relative to our primary standards. B-V and V-R colors consistent with the Sun (as measured by Landolt 1983) were used to select the secondary solar analogs. Catalog numbers for these stars within the Landolt (1983) reference are: 93–101, 98–978, 102–1081, 105–56, 107–684, 107–998, 110–361, 112–1333, 113–276, 115–271.

Apart from needing to use solar analog stars (as proxies for the Sun itself), we did follow the exact same spectrograph measurement procedures for each object (whether an asteroid or solar analog star). Specifically, asteroids and solar analog spectral images were recorded through the identical optical paths by acquiring them through the telescope and positioning them for measurement in the exact same manner within the instrumentation. Our procedures included endeavoring to produce our spectral images consistently on exactly the same detector pixels, thereby optimizing the opportunity for the mathematical division creating each reflectance spectrum to cancel out numerous potential sources of systematic errors, which can remain even after performing careful bias level and flat field calibrations.

Variable atmospheric transmission, whether arising from measuring objects at different altitudes in the sky (creating different path lengths for the light through the atmosphere, described as the air mass) or through actual changes in clear sky water vapor abundance is an additional complication. In our survey, we sought to minimize atmospheric effects in multiple ways. First, as much as possible all measurements of our NEO targets and standard stars were observed near the meridian at their highest altitude (minimum air mass) and their minimum atmospheric dispersion, where the parallactic angle for atmospheric dispersion coincides with the fixed north-south alignment of the long axis for our 0.8 × 15 arcsec spectroscopic slit. Second, we calculated our reflectance spectra ratios using the solar analog stars measured most closely in time and airmass as each target object. We additionally account for Earth atmospheric effects by noting that during spectral exposures of each object's flux through the slit and grating onto the detector, background flux from the night sky was also simultaneously recorded. Subtracting the unwanted sky flux was achieved by alternating spectral image frames between two different positions (usually noted as the A and B positions) on the slit, with 120 s being the typical exposure duration at each slit location. (Longer exposures become increasingly problematic owing to the accumulation of cosmic ray hits on the detector.) Subtracting the image frames A-B and B-A effectively (but not perfectly) removes the sky flux that was present at the same detector pixel locations as the object spectrum. All spectral images were thus obtained in pairs, with two to three sets of eight images being taken for each object. The total integration time for each of these target objects typically ranged from 30 to 120 min. Solar analog stars were measured in an identical fashion at the same A and B locations on the slit. These positions were held fixed by automated active guiding software named "Guide Dog." By active guiding to maintain these fixed locations, we achieved the maximum assurance of all measured flux being dispersed across the identical rows and pixels of the imaging CCD.

Reduction of spectral images was performed using a combination of routines within the Image Reduction and Analysis Facility (IRAF) and Interactive Data Language (IDL). We developed a software tool called "autospex" to streamline the reduction procedures, which works by automatically writing macros containing sets of IRAF (or IDL) command files. We executed our autospex procedures on single night datasets at a time, where intermediate products in the pipeline were inspected by the user at each stage. These stages include: trimming the images down to their useful area, creating a bad pixel map from flat field images, flat field correcting all images, performing the sky subtraction between AB image pairs, registering the spectra in both the wavelength and spatial dimensions, co-adding the spectral images for individual objects, extracting the 2-D spectra from co-added images, and then applying the final wavelength calibration. IDL was used for a final atmospheric correction for any difference in the air mass between the target object and its solar analog. This final correction was accomplished using the atmospheric transmission (ATRAN) model by Lord (1992) by determining a coefficient for each object and star pair that best minimizes atmospheric water absorption effects for that pair. This coefficient correction is most important near 1.4- and 2.0- microns, locations of major absorption bands due to telluric $H_2O$. The final IDL step averages all the object and standard pairs to create the final reflectance spectrum for each object. Given the orders of magnitude difference in actual flux collected for bright analog stars relative to faint asteroids, we follow the tradition to normalize the final reflectance spectrum to an arbitrary wavelength. Historically for visible wavelength spectra, this normalization to unity is made at 0.55-microns, the center of the V filter band. For our SpeX near-infrared spectra that are not linked with overlapping visible spectra, we choose the normalization wavelength to be 1.215 μm, the center of the J filter band.

2.1. Building the full data set with visible spectra and external data

While our contribution of more than 750 new near-infrared spectra represents a large ground-based telescopic contribution to NEO physical studies, the most robust scientific analysis of the population requires incorporating these new data into the context of existing NEO color measurements. For illustrating this context and making clear all of the types of color information that can be folded together, Fig. 2 presents a schematic visualization of progressively diagnostic data products. The same asteroid, 433 Eros, is depicted in each case. We follow the Fig. 2 progression in building our data set, that all totaled yields more than 1400 individual objects in Appendix II.

We discuss the progression shown in Fig. 2 briefly. Historically, the first NEO color measurements were broad-band filter photometry (Millis et al., 1976; Bowell et al., 1978), but quickly progressed to

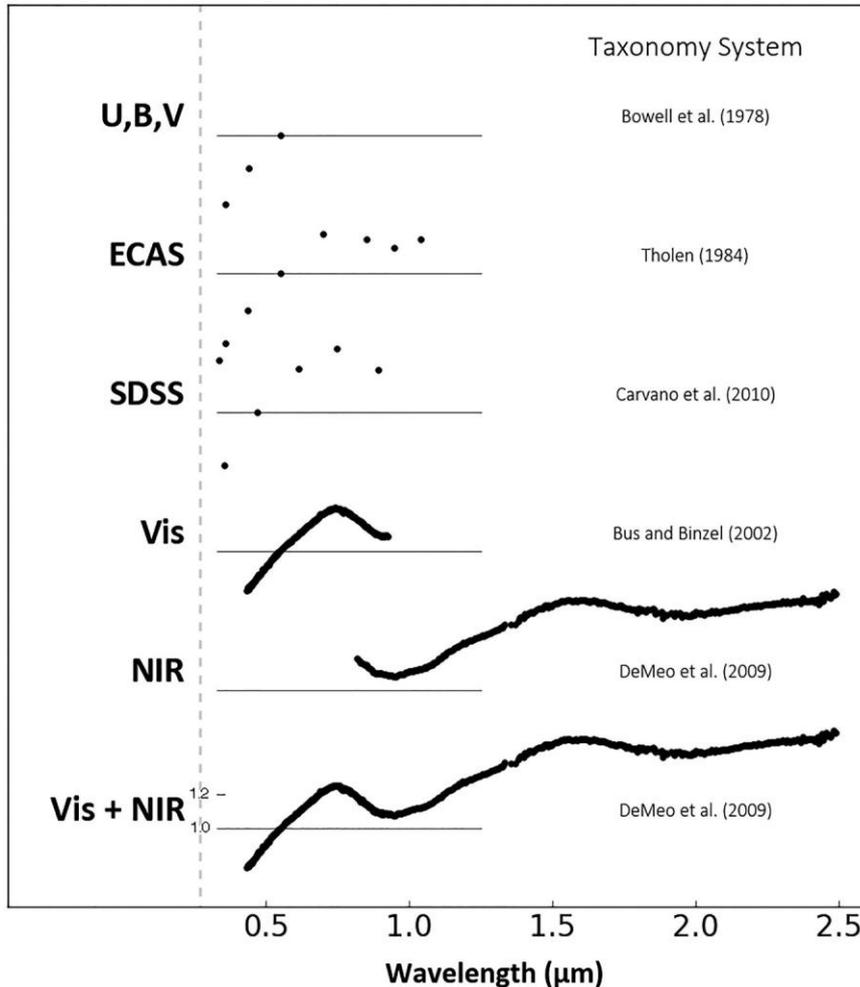

**Fig. 2.** Historical progression in telescopic measurements of asteroid spectral colors and the taxonomic systems applied to their classification. Near-Earth asteroid 433 Eros is illustrated in each case, consistently assigned to the "Sclass" by all classification methods. (The original UBV measurements were reported by Millis et al., 1976.) Our statistical analysis of the near-Earth object population follows the progression shown here, utilizing the most extensive data type available for making an object's classification assignment. The vertical scale (the same for each data set, scale shown at bottom) depicts relative reflectance ratioed to the Sun, normalized to unity at 0.55 μm (horizontal bar). See Section 2.1 for a full description.

multi-filter photometry such as the Eight Color Asteroid Survey (ECAS; Tholen 1984, Zellner et al., 1985) revealing the presence of spectral absorption bands. All sky surveys such as the Sloan Digital Sky Survey (SDSS; Ivezic et al., 2001; Carvano et al., 2010) employ similar multifilter photometry methods for efficient color reconnaissance of enormous numbers of sources, yielding an abundance of measurements for asteroids (e.g. Carry et al., 2016). Much more detailed spectral signatures are mapped out by visible wavelength spectrographs (e.g. Bus and Binzel 2002a; herein we denote visible wavelengths by VIS) and near-infrared spectrographs (denoted by NIR; e.g. DeMeo et al., 2009). Merging together both visible and near-infrared spectra (VIS

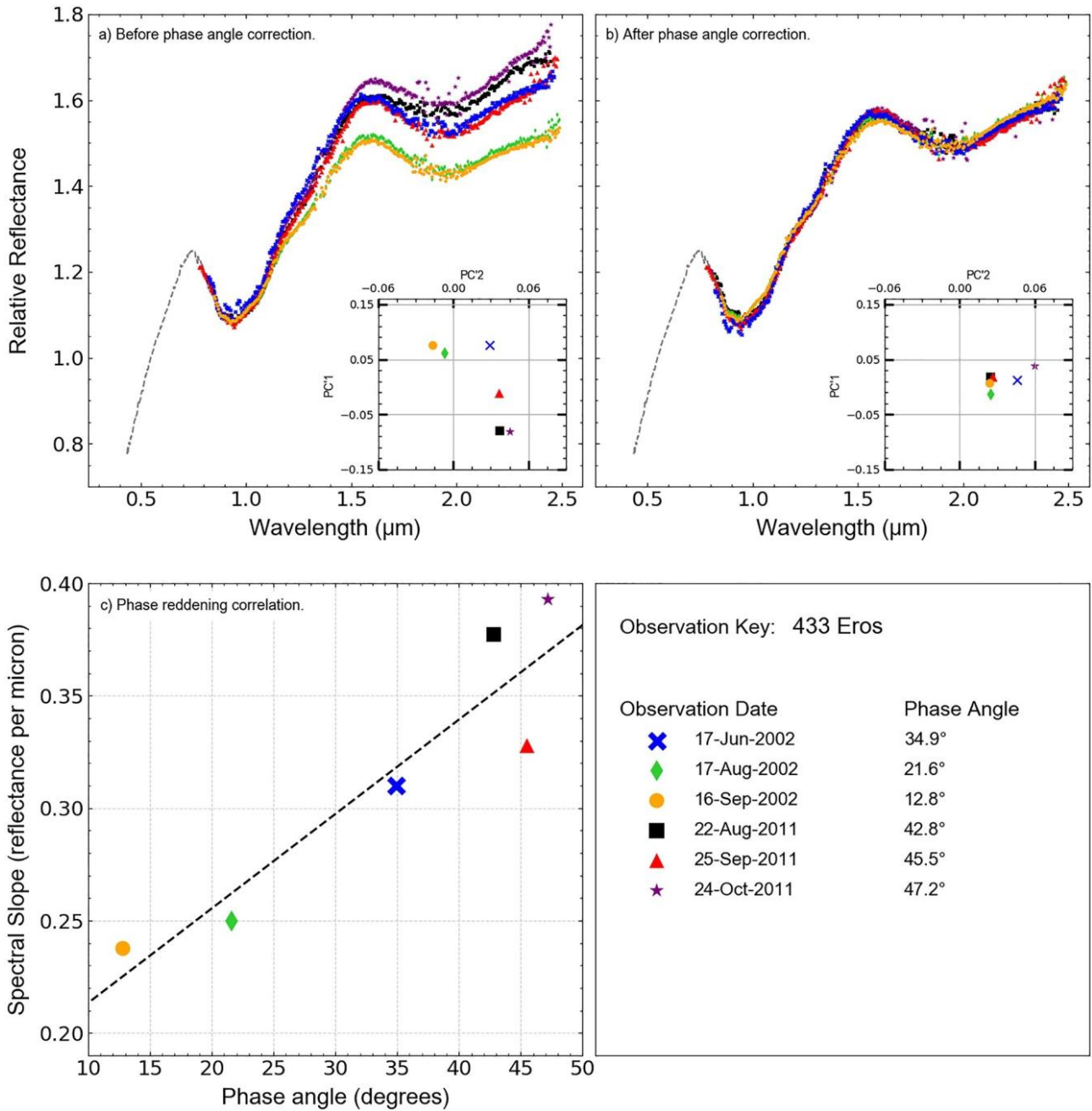

**Fig. 3.** Examination of observational systematic effects in near-infrared asteroid spectral data. **(a)** Overlay of MITHNEOS spectra of 433 Eros, measured and reduced independently on six nights separated across multiple years as indicated in the Observation Key. All spectra are normalized to unity at 0.55 μm using an average for the visible spectrum (dashed line; Bus and Binzel 2002a; Binzel et al., 2004). Their dispersion in principal component space is shown in the inset figure and discussed in Section 3.1. Most of the dispersion is accounted for by a spectral reddening correlation with phase angle, best fit **(panel c)** by an increase in spectral slope corresponding to a coefficient of +0.0036 reflectance units per degree of phase angle change. Applying this coefficient and correcting all spectra to a common phase angle of 30° shows minimal residual systematic effects in repeated MITHNEOS independent measurements **(panel b)** and greatly reduced dispersion in principal component space (inset).

+NIR) illustrates the advantage that can be gained by spanning the broadest possible wavelength range. Outcomes of this full wavelength range include advanced taxonomic classification (Section 3.3) and mineralogical modeling and possible meteorite correlations (Section 5).

Incorporating all of the data products depicted in Fig. 2 allows us to expand our entries in Appendix II to more than 1400 objects. To our set of 750 new observations presented here, we are able to add filter photometry information (sufficient for the taxonomic analysis in Section 4) for 300 objects, as deduced from SDSS colors (Carvano et al., 2010; Hasselmann et al., 2012; Ivezic et al., 2001; Solano et al., 2013; Carry et al., 2016). Here we follow the implementation described by DeMeo and Carry (2013). Visible wavelength spectra enable the compilation of an additional 300 objects into the taxonomic analysis, where these spectra are primarily derived from Bus and Binzel (2002a), Lazzarin et al. (2005) and Hicks et al. (1998), and many other sources. Approximately 100 additional near-infrared spectra are derived from Vernazza (2006), de Leon et al. (2010) and Thomas et al. (2014), and more. We emphasize that the tabulation of Appendix II carries more than 100 individual source references for the added entries. Thus, users who quote a result for any specific object (or small subset) are enjoined to cite properly the original data reference(s) and not this compilation.

An overarching objective of our compilation is to assemble full wavelength coverage (visible plus near-infrared, spanning 0.45–2.45 µm) spectra for as many objects as possible. Over this wavelength range, mineralogical interpretation becomes viable (e.g. Gaffey et al., 1993). All totaled, we are able to achieve VIS+NIR wavelength coverage for 332 objects, giving a large enough sample for robust analysis that extends beyond the labels of asteroid taxonomy. We follow this approach in Section 5.

**3. Systematic effects, corrections, and taxonomic classification**

*3.1. Phase angle effects*

For unveiling systematic effects and the limits of reliability in our survey, we took advantage of asteroid 433 Eros as a test case. For Eros, we obtained six independent spectral measurements over the course of a decade. These measurements spanned a phase angle range from 10 to 50°. (The phase angle is the Earth-Sun separation angle, as seen from the asteroid.) Eros has the advantage of being verified by in situ imaging as having highly homogeneous spectral colors over hemispheric scales (McFadden et al., 2001; Izenberg et al., 2003). Thus any date-to-date variations in our own telescopic measurements of Eros reveal the severity of possible systematic effects induced by limitations in our observation and reduction techniques. Most specifically, systematic effects may arise from night-to-night variations in atmospheric conditions and their correction, variations in the quality of CCD flat field corrections or wavelength calibrations, variations in instrument set up or observational procedures (alignment of spectrograph slit and gratings; telescope tracking accuracy; standard star selections, etc.). While cognizant of these effects and how they may be controlled, the outcomes of residual systematic effects (or "systematic errors" ) are best examined by the repeatability of individual measurements. In fact, Fig. 3a reveals that our multitude of Eros measurements show a significant divergence in slope among the individual spectra. (As noted in the figure caption, we use a single average visible spectrum so that all slope variations are uniformly leveraged by the same near-infrared spectral range.) Broadly viewed, however, the Eros spectra give a relatively consistent presentation of absorption band depths and overall shapes.

As a first step in exploring the divergence present in our Eros spectral data set, we were guided by the findings of
Sanchez et al. (2012) who convincingly demonstrated the reality of phase angle effects on spectral slope and by Thomas et al. (2014) who noted band area measurements showed a phase angle effect. We examined our Eros dataset for a similar correlation (Fig. 3c), where each calculated value for spectral slope (change in normalized reflectance per micron) follows the methodology of the Bus-DeMeo taxonomy (DeMeo et al., 2009). In agreement with Sanchez et al. (2012), we find a strong correlation between spectral slope and phase angle ($r = 0.94$; confidence level > 99%) corresponding to a +0.0036 ( ± 0.0012) increase in relative reflectance per degree. (The units are reflectance micron$^{-1}$ deg$^{-1}$, where reflectance is normalized to unity at 0.55 µm.) We also verified our findings for phase reddening of asteroid 433 Eros relative to the in situ values measured by the Near-Earth Asteroid Rendezvous (NEAR) mission. This validation required comparing two specific wavelengths from our higher resolution spectra corresponding to the NEAR camera filters. We chose for comparison the 2.3/0.95 µm ratio whose results are presented by Veverka et al. (2000). From our spectra we find the 2.3/0.95 µm ratio yields a phase reddening slope increase of +0.005 per degree that is in comparably good agreement with the Veverka et al. (2000; see their Fig. 14B) value of +0.006 per degree.

We apply our derived +0.0036 correction coefficient to adjust the spectral slopes for our Eros data, choosing to make all corrections to the spectral slope that would be observed at a common phase angle of 30°. We correct our Eros data to 30° (rather than 0°) for several reasons, including: 30° is effectively the average phase angle for our Eros data set, thus keeping each individual correction as small as possible; 30° well represents the average phase angle for all NEO spectral measurements; 30° corresponds to the phase angle for most laboratory meteorite measurements.

Fig. 3b shows the results of our phase angle correction for Eros, effectively eliminating the divergence present in the original spectra. The high level of agreement of all six spectra taken independently over a decade gives confidence towards minimal instrumental systematic errors being present within our overall data set. Detailed examination of how each individual spectrum maps out the 1- and 2-micron absorption bands reveals the limits of perfect repeatability owing to random noise and systematic variations inherent in night-to-night observing conditions.

Our long-term program conducted repeated observations of three additional S-type objects, and for each of these we performed an identical spectral slope versus phase angle analysis. These objects and their phase correction coefficients [in brackets] were: 1036 Ganymed [0.0031], 1627 Ivar [0.0020], and 4179 Toutatis [0.0028]; each with an estimated uncertainty of ± 0.0012 as found for the Eros data. A similar range of values is also found for V-type asteroids (Reddy et al., 2012a). Our conservative finding from this multiple object investigation is that there is no single phase correction that is applicable to all objects; the value may be dependent on the individual object. (Also, we have no measure for the coefficients applicable to C- or X-type objects.) For this reason, our computed and tabulated principal component scores (see Section 3.3) are based on the objects as measured at the observational circumstances reported in Appendix II. Thus an inherent scatter of about ± 0.05 in PC1' and ± 0.03 in PC2' must be recognized as being present in these data, as seen in the inset plot of Fig. 3a. In Section 4.1 we show that this uncertainty is small compared to the broad domains of the taxonomic classes. We also investigate quantitatively the relative uncertainty this may impose on our mineralogical and meteorite analysis in Section 5.

*3.2. Thermal tails*

Equilibrium temperatures for (atmosphereless) planetary surfaces near 1 AU can reach 300 K for the case of a highly absorbing (low albedo) body. The standard thermal model for asteroids (Lebofsky and Spencer 1989) shows that these bodies can emit a significant amount of flux in the short wavelength tail of their black body curve at wavelengths well below 2.5 µm. This near-infrared thermal flux is readily detectable by modern spectrometers such as SpeX (Rayner et al., 2003) as reported by Abell (2003) and detected frequently among MITHNEOS objects. Templates for deducing an albedo value from these "thermal tails" (Fig. 4) are presented by Rivkin et al. (2005) and also explored by Reddy et al. (2012b).

Nearly 50 objects in the sample we report here (including two comets) show evidence of thermal tails, thus allowing us to derive model values for their albedos. Table 1 presents the results of our fitting these thermal tails, where in some cases the fit yields a null result (listed as "no thermal tail" in the table). These "null result" cases are spectra having curvature near 2.5 μm but this curvature is not well fit by a thermal model. The spectral curvature may be intrinsic (or some systematic effect), but since a thermal fit cannot be made no albedo estimate is achieved in these cases. To fit for the cases where thermal flux is present, we apply the near-Earth asteroid thermal model (NEATM) developed by Harris (1998) using a model by Volquardsen et al. (2007) and implemented by Moskovitz et al. (2017a). This code takes several input values including the slope parameter G, surface emissivity, absolute magnitude H, geocentric and heliocentric distances at the time of observation, and the phase angle at observation. The slope parameter G = 0.05 is assumed for all objects, which is a reasonable average our *a priori* assumption that in all model cases we are dealing with low albedo surfaces (Lagerkvist and Magnusson, 1990). As is common practice when using NEATM, the surface emissivity is assumed to be 0.9. The absolute magnitude and orbital information were retrieved from the JPL Horizons system. NEATM also requires a thermal infrared beaming parameter η, which we compute from an empirical relationship between η and the phase angle at the time of observation (Masiero et al., 2011). Holding this value of η fixed, the model returns best fits to the geometric albedo and object diameter. This specific implementation of the NEATM is not directly sensitive to object diameter because we are modeling thermal excess relative to a normalized near-IR reflectance; therefore we do not tabulate a diameter estimate.

The fitting procedure begins by computing the reflectance ratio between the visible (0.55 μm, reference wavelength for absolute magnitude) and near-IR (at 2.5 μm). This necessitates estimating the reflectance without any thermal contribution at 2.5 μm. This is estimated by extrapolation of a linear fit to the spectrum between 1.5 and 1.9 μm, a procedure consistent with previous works (e.g. Rivkin et al., 2005, Reddy et al., 2012a). The reflectance at 0.55 μm is directly measured for objects with a visible spectrum or is estimated by extrapolation of a linear fit to the spectrum at wavelengths < 1 μm.

The model scans through a range of albedo from 0.001 to 0.15, subtracts a NEATM model for each albedo, and calculates residuals between the thermally corrected spectra and an extrapolation of the linear fit to the data between 1.5 and 1.9 μm. The best-fit albedo is determined based on minimized residuals. Fig. 4 shows an example of our NEATM fit and the resulting spectrum of the asteroid after removal of the modeled thermal flux. We infer the flattened thermal flux subtracted spectrum to be representative of the intrinsic reflectance properties of the surface material for our subsequent analysis.

We do not analyze albedo data further, noting that much more complete data sets are available. For example, see Masiero et al. (2017). *3.3. Taxonomic classification tabulation*

The progression of data types illustrated in Fig. 2 is matched by the progression of methodologies for asteroid taxonomy. The data constrain the number of free parameters from which the taxonomic system can be defined. In this work our first objective is to assign, as robustly as possible, a taxonomic class to each entry in Appendix II. The wavelength coverage and the spectral resolution of the measurements available for each entry determine which taxonomic classification scheme we apply, and we do so following the progression detailed in Fig. 2, up to the limit of the available data. In other words, if only filter photometry data are available for a particular object, we tabulate the taxonomic class respectively defined by Tholen (1984) for ECAS data; for SDSS data we follow the implementation method of DeMeo and Carry (2013). If more extensive data are available, we move down the progression applying the Bus taxonomy (Bus 1999; Bus and Binzel 2002b) when visible spectra are available or the near-infrared only taxonomy of DeMeo et al. (2009). As is illustrated in the case of 433 Eros (Fig. 2), when the full suite of data products is available we apply the farthest progression in the taxonomic classification scheme. Thus in Appendix II, for 433 Eros we tabulate the results of the Bus-DeMeo taxonomy analysis (DeMeo et al., 2009) that takes into account the full availability of visible plus near-infrared (VIS+NIR) spectral information over the 0.45–2.45 μm wavelength range. We note that each of the advancing taxonomic systems was developed with a backward compatible philosophy; for example 433 Eros is an S-class asteroid in all systems. To be sure, advancing taxonomic systems found distinguishing characteristics to justify new classes and subclasses, but with clear traceability to prior categorizations. DeMeo et al. (2015; see their Table 1) gives a detailed synopsis.

Our Appendix II tabulation of the Bus-DeMeo taxonomy results was accomplished using the online classification tool developed by Stephen M. Slivan (Wellesley College) and publicly available at smass.mit.edu. Most specifically we applied "Version 4" of the Bus-DeMeo classification system (implemented August 2013) that removed ambiguities in distinguishing O-type, B-type, and C, Ch, Xk-type based on high order principal component scores. The classification tool at smass.mit.edu gives full details of the flow chart used in implementing Bus-DeMeo taxonomy, maintaining full consistency with the class definitions and distinctions proposed by DeMeo et al. (2009). Sufficient data for a new class, dubbed "Xn" is described below. Interestingly, the need for these refinements became apparent in processing the large volume of new data from this survey, where the NEO population often shows end members and spectral characteristics that are more pronounced than main-belt asteroids that constitute the bulk of the basis set defining all taxonomic systems.

Corrections to our spectral data were applied prior to taxonomic classification only for the case of thermal tails (Section 3.2), i.e. thermal tails were removed prior to executing our classification routines. We note that the Bus taxonomy and the Bus-DeMeo taxonomy calculate and remove the spectral slope prior to calculating principal component scores, thus reddening effects due to phase angle and space weathering have minimal consequence to the classification outcome. Systematic effects do remain owing to night-to-night variability in observing conditions for Earth-based telescopes; the insets of principal component scores in Fig. 3 illustrate this systematic range. All systematic variability is small compared to the domains of broad taxonomic classes (see Section 3.3.2), and comes into play only in the cases of classification very near to class boundaries. For all cases of the classes we tabulate in Appendix II, a visual inspection of the spectrum and the algorithm's class assignment was made to ensure no potential systematic effects pushed a classification across a boundary in a way that was not judged to be sensible. As detailed in DeMeo et al., (2009), this visual inspection is the last step in assigning a taxonomic class.

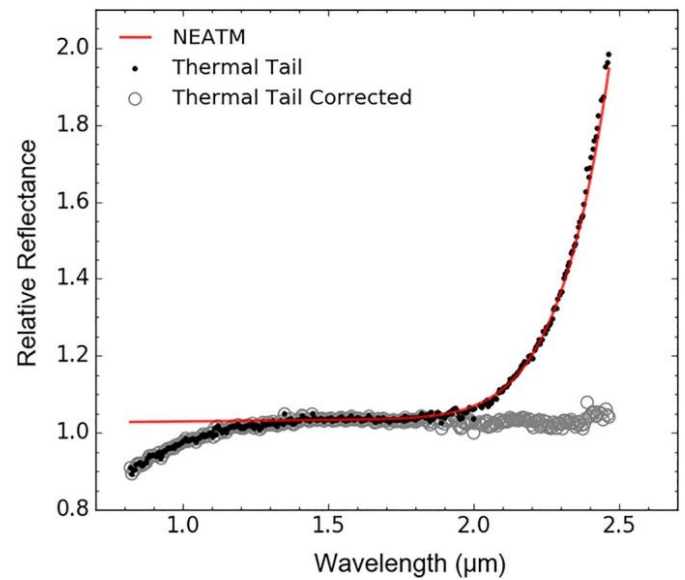

**Fig. 4.** Example MITHNEOS spectrum displaying a thermal tail (solid points); 2012 LZ1 measured on 14-June-2012 UT at 1.05 AU from the Sun. A solid line shows the best fit Near-Earth asteroid thermal model (NEATM; Harris 1998) based on the implementation by Moskovitz et al. (2017a) for an albedo = 0.017 and thermal parameter η = 1.094. The open circles depict the spectrum with the modeled thermal flux removed. Taxonomic classification (C in this case) is derived from the thermally corrected spectrum.

**Table 1**
Tabulation of thermal modeling results for near-Earth objects displaying strong enhanced flux near 2.5 μm, arising from the warm temperatures for low albedo surfaces at heliocentric distances near 1 AU. The NEATM thermal model (Harris 1998) was successfully fit to 45 out of 47 cases we investigated. Two of the objects measured are comets. The values we report here are geometric albedos and the beaming parameter η, as described in Section 3.2.

| Number | Name | Provisional Designation | NEATM Parameters | | | Notes |
|---|---|---|---|---|---|---|
| | | | Albedo | Radius | Eta | |
| 1580 | Betulia | 1950 KA | 0.079 | 2.976 | 1.373 | |

| Number | Name | Designation | | | | Notes |
|---|---|---|---|---|---|---|
| 3552 | Don Quixote | 1983 SA | 0.004 | 27.635 | 1.311 | Poor fit |
| 7753 | | 1988 XB | 0.067 | 0.489 | 1.292 | |
| 11885 | Summanus | 1990 SS | 0.027 | 0.807 | 1.184 | |
| 17274 | | 2000 LC16 | 0.025 | 2.012 | 0.958 | |
| 26760 | | 2001 KP41 | 0.020 | 4.092 | 1.149 | |
| 52762 | | 1998 MT24 | 0.021 | 5.028 | 1.241 | |
| 53319 | | 1999 JM8 | 0.010 | 6.060 | 1.410 | |
| 65996 | | 1998 MX5 | 0.010 | 1.326 | 1.165 | |
| 85774 | | 1998 UT18 | 0.025 | 0.607 | 1.223 | |
| 90416 | | 2003 YK118 | 0.008 | 1.482 | 1.207 | Poor fit |
| 108519 | | 2001 LF | – | – | – | No thermal tail (A > 0.15) |
| 139359 | | 2001 ME1 | 0.026 | 1.884 | 1.030 | |
| 153219 | | 2000 YM29 | 0.038 | 0.712 | 0.990 | |
| 153591 | | 2001 SN263 | 0.126 | 0.780 | 1.083 | Triple system |
| 170502 | | 2003 WM7 | 0.124 | 0.685 | 1.019 | |
| 170891 | | 2004 TY16 | 0.025 | 1.673 | 1.033 | Poor fit |
| 175706 | | 1996 FG3 | 0.082 | 0.485 | 0.880 | |
| 234061 | | 1999 HE1 | 0.022 | 1.353 | 0.972 | |
| 256412 | | 2007 BT2 | 0.042 | 1.233 | 1.164 | |
| 275792 | | 2001 QH142 | 0.019 | 1.007 | 1.014 | |
| 285263 | | 1998 QE2 | 0.150 | 0.821 | 1.001 | |
| 307005 | | 2001 XP1 | 0.011 | 1.591 | 1.450 | |
| 308635 | | 2005 YU55 | 0.050 | 0.124 | 1.021 | |
| 326732 | | 2003 HB6 | 0.018 | 1.496 | 1.120 | |
| 329291 | | 2000 JB6 | 0.071 | 0.360 | 1.059 | |
| 354030 | | 2001 RB1 | 0.033 | 0.730 | 0.968 | Poor fit |
| 363505 | | 2003 UC20 | 0.020 | 1.127 | 1.514 | |
| 438105 | | 2005 GO22 | 0.042 | 0.618 | 1.047 | |
| 451124 | | 2009 KC3 | 0.015 | 1.363 | 1.352 | |
| 452561 | | 2005 AB | 0.039 | 1.064 | 1.324 | |
| 455322 | | 2002 NX18 | 0.012 | 2.103 | 1.469 | |
| 475665 | | 2006 VY13 | 0.026 | 1.496 | 1.259 | |
| 481032 | | 2004 YZ23 | 0.081 | 2.230 | 0.948 | |
| | NEAT | 2002 EX12 | 0.009 | 4.628 | 1.148 | Comet 169P |
| | | 2004 QD3 | – | – | – | No thermal tail (A > 0.15) |
| | Siding Spring | 2004 TU12 | 0.008 | 10.739 | 1.326 | Comet 162P |
| | | 2005 GR33 | 0.040 | 0.132 | 1.300 | |
| | | 2005 JB | 0.011 | 1.101 | 1.370 | |
| | | 2006 RZ | 0.045 | 0.273 | 1.161 | |
| | | 2007 PF28 | 0.051 | 0.406 | 1.225 | |
| | | 2008 SV11 | 0.067 | 0.536 | 1.864 | |
| | | 2008 US4 | 0.134 | 0.120 | 0.922 | |
| | | 2011 BE38 | 0.004 | 2.195 | 1.501 | Poor fit |
| | | 2011 SR5 | 0.023 | 0.317 | 1.442 | |
| | | 2012 LZ1 | 0.017 | 0.534 | 1.094 | |
| | | 2012 RM15 | 0.077 | 0.030 | 0.928 | |

### 3.3.1. New Xn class

Six spectra in our NEO survey show a unique narrow absorption feature centered at 0.9 μm, not previously categorized in the "featurebased" Bus-DeMeo system. Asteroid 3671 Dionysus is the lowest numbered NEO in our data set displaying this feature. Notably, such a relatively flat overall (0.45–2.45 μm) spectrum containing a narrow 0.9micron feature was previously unique to the main-belt asteroid 44 Nysa. Thus "Version 4" of the Bus-DeMeo classification tool (discussed above) allows for input spectra to receive the Xn class as a viable taxonomic assignment.

### 3.3.2. Quantitative use of principal component analysis

Our use of principal component (PC) scores for quantifying residual systematic effects in our data (Section 3.1), and in our overall population analysis (Section 4), merits a brief recap of their genesis and interpretation that goes beyond their role in assigning taxonomic classes. The mathematical technique of principal component analysis (PCA) was introduced into the asteroid field by Tholen (1984) and has been an ongoing foundation for the evolution of asteroid taxonomy. Tholen and Barucci (1989) and Bus (1999) give excellent explanations (and further references) of PCA mathematics applied to asteroid spectroscopy. In the example of the Bus-DeMeo taxonomy, a sample of 341 asteroid spectra were binned over 40 equally spaced wavelengths spanning the range 0.45–2.45 μm. Principal component analysis was used to find the vector through this 341 × 40 dimensional space along which exists the greatest variance. This vector of greatest variance, comprised by 40 coefficients (one for each wavelength), is known as the first eigenvector. For any object spectrum, multiplying its measured reflectance times the eigenvector coefficient at that wavelength, and summing the 40 values, yields the first principal component score. While traditionally this resulting number would be denoted as the PC1 score, the Bus-DeMeo taxonomy uses the notation PC1' ("P C one prime") to distinctly signify that spectral slope is removed prior to applying the PCA methodology. The second eigenvector (from which the PC2' score is calculated) is found to be the vector having the maximum variance *orthogonal* to the first eigenvector. A total of 39 additional eigenvectors can be found where each is orthogonal to all previously calculated. Each progressive eigenvector, however, accounts for less and less of the total variance of the input data set.

The power of principal component analysis is the reduction of an enormous multi-dimensional space in to a way that highlights how the individual members of the data set are maximally different from one another. Thus a plot of PC2' vs. PC1' displays the two dimensions of greatest variance of asteroid spectral properties. Because slope is removed, these two principal components are most sensitively giving quantitative information on the spectral features. Quantitatively, DeMeo et al. (2009) finds that PC1' accounts for 63% of the variance and is a measure of the 1-micron band characteristics while PC2' (24% of the variance) is a measure of the 2-micron band characteristics. This fact allows us to interpret the residual scatter of our repeated 433 Eros spectral measurements in Fig. 3. The inset of Fig. 3b shows that most of the residual scatter is along the dimension of PC2', representing our ability to map the structure of the 2-micron absorption band. Quantitatively we see that residual atmospheric effects, most notably water vapor absorption near 2-microns, provides the limiting factor in the repeatability of our measurements.

We emphasize here, for use in Section 8, the importance of spectral analysis that is independent of spectral slope for the reasons outlined in Section 3.1. For any analysis that uses Bus-DeMeo principal component scores, that analysis is "slope free" because as noted above, slope is removed prior to the computation of PC1', PC2', etc. Thus any measured spectral alteration requires actual changes in spectral features and is much less influenced by secondary effects related to particle size or phase angle effects.

**4. Population distributions and spectral alteration effects**

We begin our analysis of the near-Earth object population by examining distributions of taxonomic types among various orbital subsets. We then advance to quantitative investigation of spectral properties using principal component analysis, in particular, we explore the trends in spectral characteristics that arise from alteration effects such as space weathering and shock darkening.

*4.1. Taxonomic distribution of the near-Earth object population*

Fig. 5 presents a histogram for the distribution of taxonomic types for more than 1000 near-Earth objects and > 300 Mars-crossers, displaying a factor of three increase in the available data from a similar depiction by Binzel et al. (2004; see their Fig. 2). A new result is finding all taxonomic classes recognized in the main-belt are now represented among the NEOs, where the Cgh class remains the most rare with one member: (365246) 2009 NE being measured by Somers et al. (2010). While the current overall spectral sample remains biased by discovery (and by corollary available for physical study) toward high albedo objects, the improvement in fainter detection limits for both discovery and physical study creates some notable differences in the observed distribution compared to the previous decade. In particular, B- and Ctypes represented only ~5% of the Binzel et al. (2004) distribution; in the current sample they are about 15%. Similarly D-types are seen in greater relative abundance among NEOs (~3% as compared to 1% previously), as recognized and interpreted by DeMeo et al. (2014). Objects in the S-complex (dominated by S- and Sq-types) are consistent at being ~50% of the sample while X-types (~10%) and V-types (~5%) also remain consistent fractions in the current sample. Q-types show an uptick of ~5–10%, which may be accountable to preferential observing of objects with extremely close encounter distances to the Earth and focus on the proposition that such encounters may result in "seismic shaking" that erases space weathering effects (Nesvorný et al., 2005, 2010; Binzel et al., 2010). YORP spin up may also play a significant role in surface refreshing (Graves et al., 2016). We discuss these processes in Section 8.

With the greater overall sample of NEO observations, subsets of the population can be examined for trends that may be insightful with regard to their source regions, at least to the extent that we can currently identify their likely escape routes from the main-belt into near-Earth space. Specific subsets of interest include objects contributing to the NEO impact hazard and the provenance of objects in the most accessible orbits for exploration and resource utilization. One can only consider this analysis as preliminary as we recognize the median diameter for our data set is ~0.7 km and only ~10% of our sample is smaller than 150 m. To start, we consider the full sample where Fig. 5 shows a very strong consistency between the overall NEO population and the subset of potentially hazardous asteroids (PHAs; compare the top two panels of Fig. 5). In fact, comparing directly the PHA versus non-PHA histogram distributions (second and third panels of Fig. 5) shows no significant difference. We also find NEOs in the most easily reached orbits (ΔV < 7 km/s) also show the same distribution. Thus we find that no taxonomic class is seen to dominate (or appear devoid) within the subsets posing the greatest long-term impact hazard to Earth or providing the most easily reached destinations. These findings are consistent with the PHA and low ΔV subsets being a dynamically mixed sample of the overall NEO population.

A dynamical distinction, however, is revealed in the panels of Fig. 5 comparing the taxonomic distributions of the near-Earth and Marscrossing populations. Specifically, we see a significantly less diverse population among the Mars crossers. Dynamically this is consistent with the Mars-crossing population being a first step in the dynamical diffusion of objects out of the inner edge of the main-belt, a population that is most naturally dominated by S-types that represent the largest fraction found in the innermost belt (DeMeo and Carry 2014). A notable contrast between the near-Earth and Mars-crossing populations is the significantly lower fraction in the Q:S asteroid ratio, interpreted as the ratio between freshly re-surfaced ordinary chondrite objects (Q) and older "undisturbed" weathered surfaces (S). The higher proportion of "fresh" Q-type surfaces among the near-Earth population is consistent with the notion of planetary encounters playing a role in resurfacing objects through seismic shaking (Nesvorný et al., 2005, 2010; Binzel the innermost main belt (DeMeo and Carry, 2014), from which the Mars-crossers are most immediately derived.

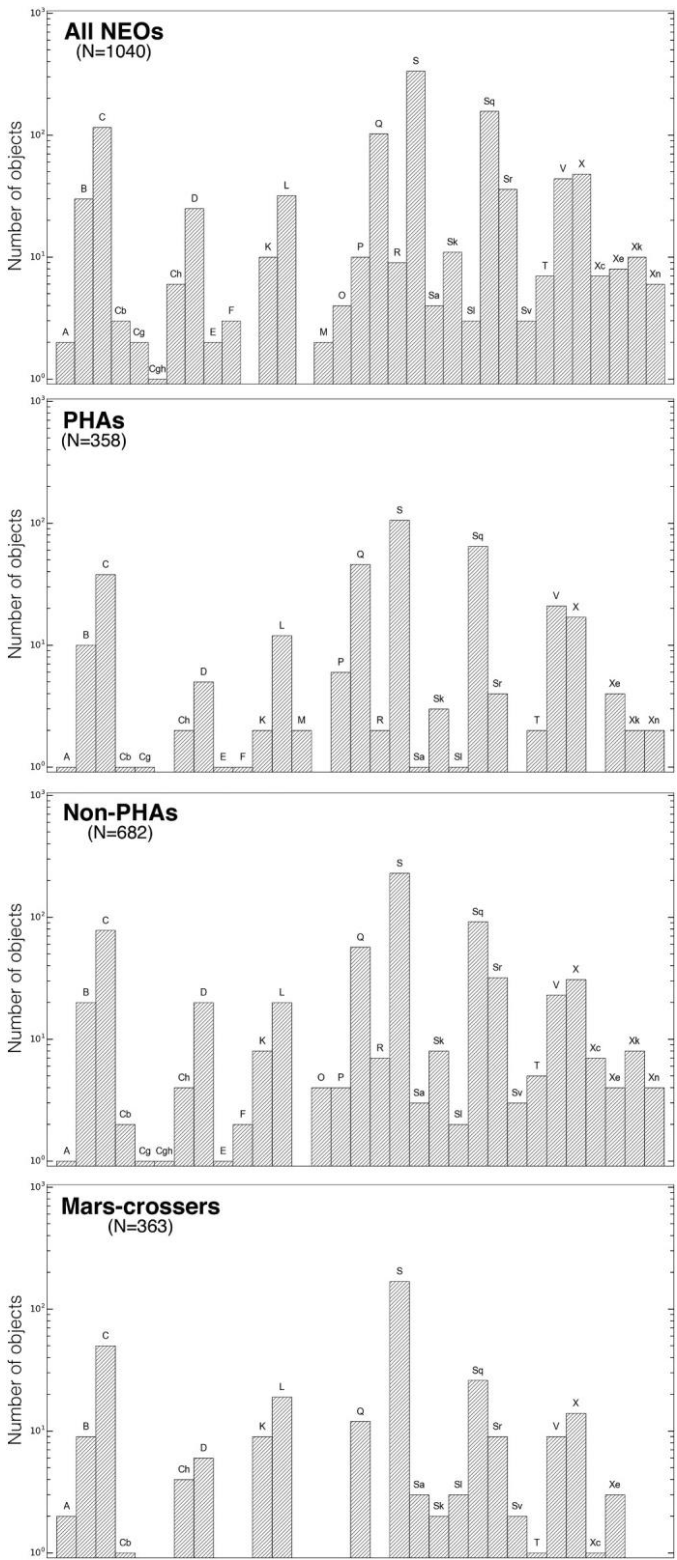

et al., 2010), though the lower perihelion values for the NEOs relative to Mars crossers also enhances YORP spin up to the point of mass shedding as a viable mechanism for exposing fresh "unweathered" subsurface material. The importance of this latter mechanism is explored by Graves et al., (2016, 2018). We examine the overall data set in terms of competing process of "resurfacing versus weathering" in Section 8.

*(caption on next page)*

**Fig. 5.** Histogram distribution for taxonomic types found in the near-Earth population and selected dynamical subsets. The full set for more than 1000 objects is shown at the top, where this full set displays the full range of taxonomic types found in the main belt, consistent with a broad distribution of main belt sources. Comparing directly the populations of Potentially Hazardous Asteroids (PHAs) with their complement of non-PHAs (second and third panels) shows no significant deviations, consistent with a high degree of dynamical mixing among the main-belt population that escapes in to the inner solar system. The sampled Mars-crossing population shows less diversity than the overall sample, and overall more closely resembles the distribution of taxonomic types found in

We next explore the mass and size dependence for the taxonomic distribution for our sample. Fig. 6 shows the proportions by mass for the major NEO taxonomic classes, where the mass calculations follow the assumptions and methodology used by DeMeo and Carry (2014). Directly comparing the NEO population with the smallest size range that can be computed for the main belt (5–20 km) shows definitively the overall concordance with the inner and middle main belt being the dominant supplier to the observed NEO population. Fig. 7 examines the diameter dependence of the major taxonomic categories using a running box mean analysis. Examining the NEO population by mass (Fig. 6) and by number of observations (Fig. 7) shows that the relative fraction of S-types remains remarkably constant over the entire size range. (The only exception is the mass dominance at the largest sizes, where the ~30 km objects 433 Eros and 1036 Ganymed are both S-types.) C-types and "other" types remain quite constant (~20% each) until sizes below 1 km are reached. In this smallest size range, C-types appear to diminish to about 10%, which if real, could be accounted for by their (assumed) lower overall strengths and depletion by collisional disruption. However, we cannot exclude observational bias effects against discovery or physical measurement of these lower albedo objects. In fact, we note that Wright et al. (2016) do find the fraction of low albedo objects to be about 25% among the discoveries in this size range achieved by the NEOWISE survey.

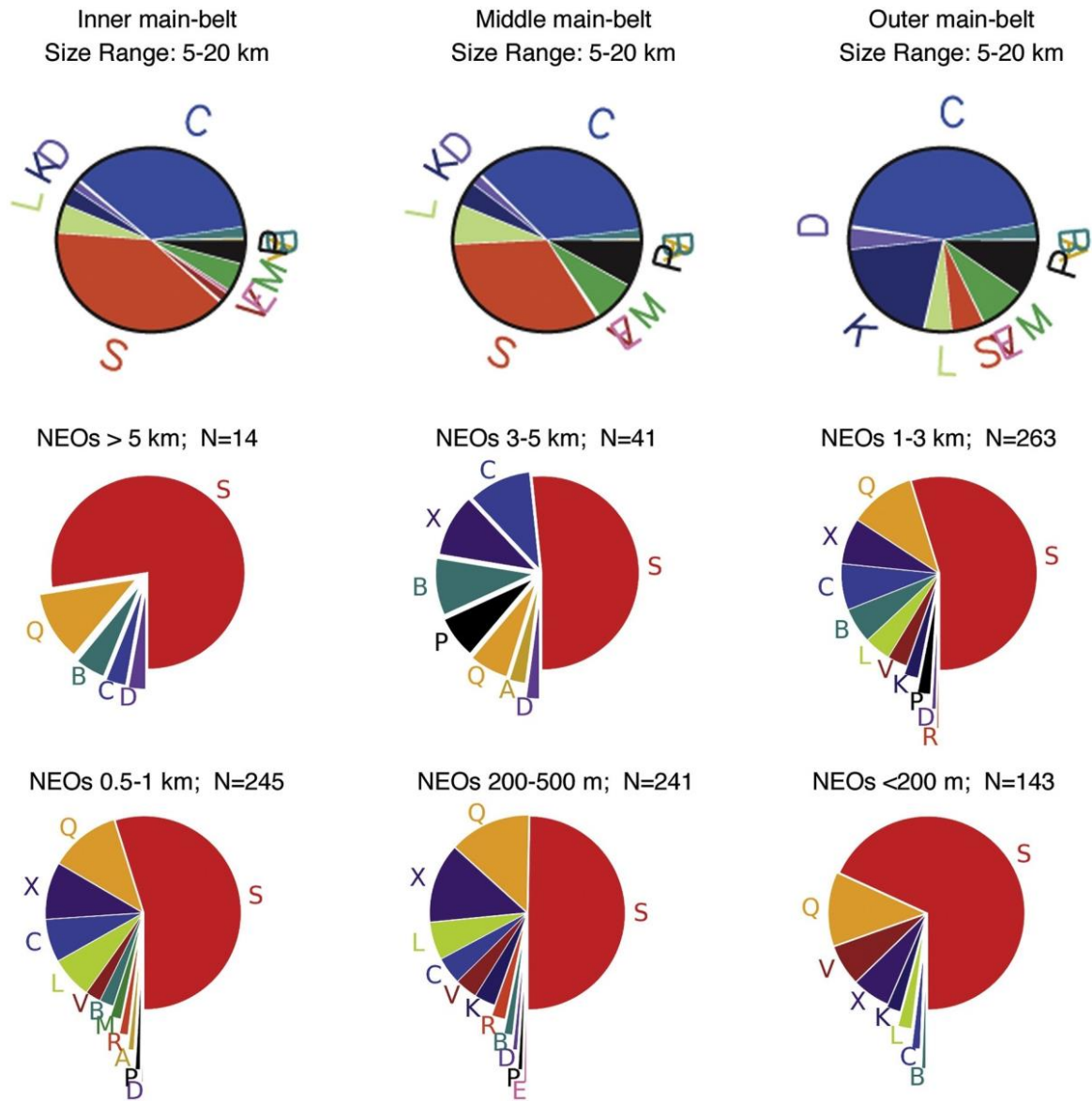

**Fig. 6.** Mass distribution of the near-Earth object population compared with the smallest size range evaluated within the main-belt. Top line depicts 5–20 km mainbelt asteroids in the inner, middle, and outer ranges. (Adapted from DeMeo and Carry 2014.) Two ~30 km S-type NEOs, 433 Eros and 1036 Ganymed, dominate the total mass in the $D > 5$ km subset. For $D < 5$ km, the mass distribution of the NEO population remains remarkably consistent although bias effects (toward discovery and characterization of higher albedo objects) likely dominate in the smaller size ranges. With the caveat that observational limits do not allow main-belt asteroid measurements in the same size range as NEOs, the concordance of the NEO taxonomic distribution with the inner and middle main belt affirms the inner half of the asteroid belt as the dominant source for the near-Earth population.

*4.2. Principal component analysis of the near-Earth object population*

While taxonomic analysis considers the NEO population in discrete bins, we also seek to explore quantitatively the continuous distribution properties of the population. As presaged by the technique summary in Section 3.3.2, we employ principal component analysis as a diagnostic tool to reveal the dimensions of our spectra having the greatest variance. Fig. 8 shows the PCA distribution for our NEO subset having full visible to nearinfrared wavelength coverage 0.45–2.45 μm, allowing us to overlay the NEO population directly on the main-belt asteroid sample used in the definition of the Bus-DeMeo taxonomy (DeMeo et al., 2009). While the overlapping regions for all major classes are clear, NEOs show a much greater overall dispersion in their spectral properties that is not easily accounted for just by plotting a larger number of objects. Among possible explanations for this dispersion, we note two that are related to intrinsic properties. (External factors, such as space weathering or shock darkening, are considered in Section 4.3.) A first intrinsic consideration is that at smaller diameters the much lower gravity (and perhaps shorter lifetimes against collisional disruption) may not favor the build up of an extensive fine-grained regolith. The overall resulting coarser surface texture of large grain sizes can produce much more distinct spectral band features as demonstrated by laboratory experiments (e.g. Cloutis et al., 2015). The more pronounced spectral bands thus produce a greater variance in PCA scores. Alternatively or perhaps concurrently at their small diameter scale, NEOs may more easily consist of pure compositional end members for mineralogies present on larger asteroids. As an explanation for this conjecture, consider that telescopic spectra measure spatially unresolved asteroids as a hemispherical average. Thus any distinct mineralogical unit (e.g. having a few hundred meter scale) on a ~100 km large asteroid is averaged together with reflected light from all other surface units; thus the distinct spectral signature of a specific unit is muted. In contrast, a distinct compositional unit comparably sized to the entirety of an individual small object (say, of order 10–100 m), can have its specific spectral signature telescopically revealed. Proximity to Earth, allowing the smallest objects to be telescopically sampled, can consequently enable distinct mineralogical units being captured in our NEO spectral data set.

*4.3. Space weathering and shock darkening trends revealed by the NEO population*

The generally younger and "fresher" surfaces of near-Earth objects have made them the focus for understanding the process of space weathering (see reviews by Binzel et al., 2015 and Brunetto et al., 2015) and possibly shock darkening (Britt and Pieters 1994). In this section we consider how these processes might be a contributing factor in the greater dispersion of spectral properties seen for the NEO population, as discussed above and evidenced in Fig. 8. Binzel et al. (2004) proposed much of the dispersion along the diagonal transition from S-types to Qtypes ("line α" in the Bus-DeMeo taxonomy; DeMeo et al., 2009) to be a size dependent correlation. The explanation offered was that NEOs at smaller sizes appear less weathered owing to their shorter collisional lifetimes (compared to large asteroids) and consequentially younger mean surface ages. Nesvorný et al. (2005) noted that rather than collisional age, close Earth encounters might be a more frequent occurrence, suggesting that tidal stresses during an extremely close approach might sufficiently disturb the passing objects to reshuffle and "refresh" their space weathered surfaces. Indeed, a correlation with surface freshness based on the minimum orbit intersection distance (MOID) was discovered observationally by Binzel et al., (2010) and analyzed further by Nesvorný et al. (2010). We discuss the interplay of these processes in Section 8.

To explore the space weathering trend among S-complex objects, we use the laboratory based model of Brunetto et al. (2006) who parameterize space weathering using the parameter Cs, which scales according to the damage measured in mineral structures resulting from ion irradiation experiments in the energy range of 60–400 keV. The Brunetto model, and its application here, has an olivine/pyroxene mineralogy as its underlying assumption. Shown as an inset within Fig. 8b is the progression of PCA scores for the spectrum of Q-type asteroid 1862 Apollo altered by increasing amounts of space weathering corresponding to values of Cs = 0.00 (initial) to Cs = 0.35 in increments of 0.05. Following that progression of alteration points effectively defines the "space weathering vector" in principal component space. Most clearly, this experiment shows how the dispersion from "fresh" asteroid surfaces to more weathered ones matches very well to the observed dispersion in asteroid spectra with a very strong parallel correlation to "line α."

Space weathering effects for low albedo C-complex objects are more

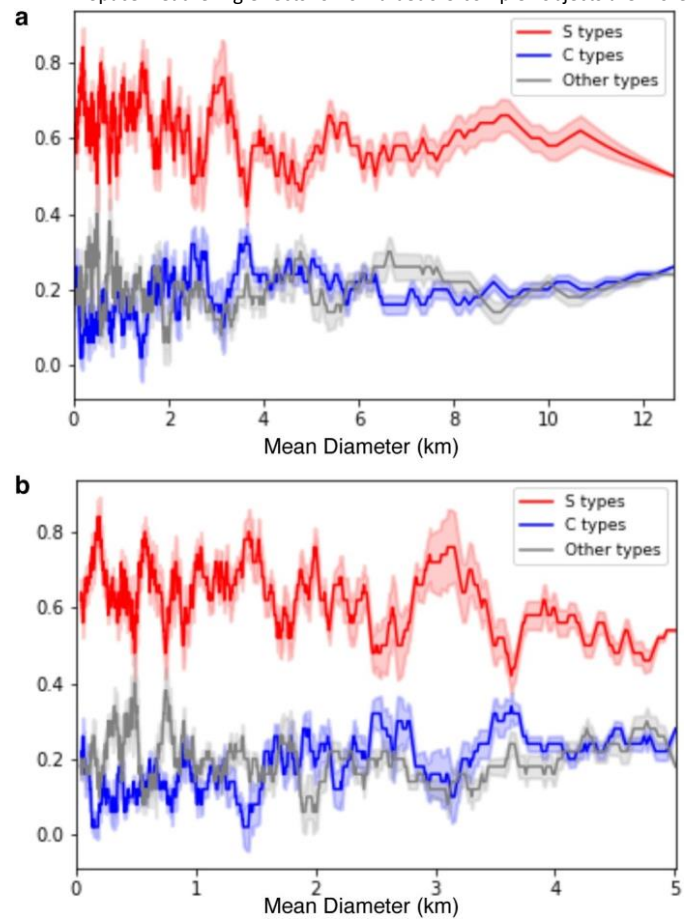

**Fig. 7.** a) Running box mean (N = 50) for the NEO population examining the diameter dependence for the distribution of taxonomic types. Objects are grouped into their major taxonomic complexes with the sum of all fractions equal to unity at any given diameter. Shaded areas depict the standard deviation of the mean. Overall the S-types maintain a remarkably constant fraction (about 60%) of the total number across the full range. Generally about 20% of the population are C-types and 20% are represented by other types. **b)** Expanded view for the size range below 5 km showing a possible trend for Ctypes dropping to about 10% below sizes of about 1.5 km. It remains an open question as to whether this change is physically real or possibly an observational bias against the detection and characterization of low albedo objects. asteroid classes.

guided by the analysis of shock properties found in the Chelyabinsk meteorite by Kohout et al. (2014), who identified two processes related to the melting of sulfides. In the first case, sulfide (troilite) melt fills the cracks within the grains, thus making them optically dark. The second case, which may be described as impact melting, occurs for higher temperatures where the silicates also melt and then recrystallize with both sulfides and iron mixed in. The corresponding peak shock pressure range for these two shock processes was estimated by Moreau et al. (2017) with a corresponding pressure range of ~40–50 GPa for sulfide melting and > 60 GPa for whole-rock melting. Reddy

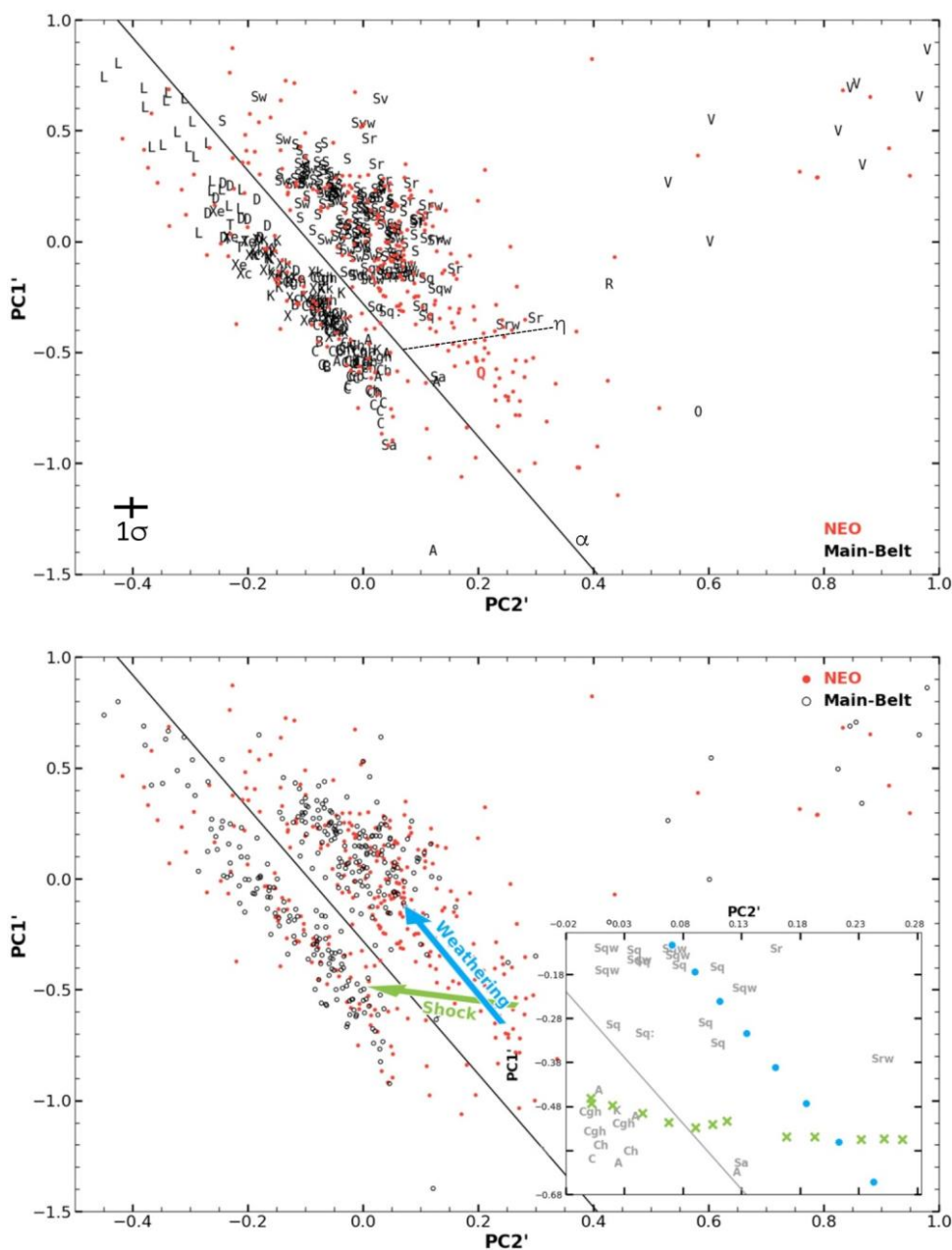

**Fig. 8.** A view of principal component space comparing the domains of main-belt and nearEarth asteroids. Plotted are the first and second principal component scores (PC1' and PC2') based on the eigenvectors defined in the BusDeMeo taxonomy (DeMeo et al., 2009), where the input data span visible plus near-infrared wavelengths and the added notation (') indicates spectral slope has been removed prior to the definition of the eigenvectors. Two boundary lines "α" and "η" important to taxonomic classification are labeled; these are discussed below and in Section 8. **Top panel:** Letters indicate the distribution of taxonomic classes for 370 main-belt asteroids defined by DeMeo et al. (2009), plus the Q-class corresponding to the near-Earth object 1862 Apollo (McFadden et al., 1985; Tholen 1984). As discussed in Section 4.2, a much greater dispersion is seen for the near-Earth objects (red, filled circles) in comparison with the mainbelt. Observational errors, described in Section 3.1 lead to uncertainties in principal component values for individual objects, depicted by the two-dimensional error bar at lower left. **Bottom panel:** Two processes are examined that may play a role in creating the dispersion of asteroid spectra as revealed in principal component space. Increasing space weathering of the Q-type asteroid 1862 Apollo (inset, filled circles; see Section 4.3 for the Brunetto et al., 2006 model description) shows that a progressive increase in space weathering can account for nearly all of the dispersion among S-complex asteroids. We depict here "the space weathering vector," which we find parallels the diagonal line in the figure corresponding to the progressive spectral dispersion for the S-complex that was revealed empirically in the asteroid data and denoted as "line α" in the (DeMeo et al., 2009) analysis. We also explored the progressive role of shock darkening, guided by the work of Reddy et al. (2014), Kohout et al. (2014, 2015) and Moreau et al. (2017), as described in Section 4.3. Within the inset figure, "x" symbols show the detailed outcomes for progressive alteration that depicts a "shock darkening vector" starting with the unshocked Chelyabinsk meteorite then adding increasing mixtures of shocked material (0, 5, 10, 20, 30, ..., 90, 95, 100%). Kohout et al. (2015) explore this progression in more detail. We note the orthogonal nature of the space weathering and the shock darkening processes, with the shock darkening vector crossing "line α" from the domain of the S-complex to the generally featureless

difficult to decode, as laboratory irradiation experiments on possibly analogous carbonaceous chondrites meteorites (such as by Lantz et al., 2017) reveals effects on both spectral slope and spectral curvature. The outcome is a divergence in spectral slopes and principal component scores that is dependent upon the underlying surface composition. The reader is referred to Lantz et al. (2018) for the detailed analysis and discussion.

To quantitatively explore the effects of shock darkening, we employ the method and results of Reddy et al. (2014) and Kohout et al. (2015). We are et al. (2014) and Kohout et al. (2015) show that these two processes yield similar outcomes in terms of their measureable reflectance spectra. Principal component scores were calculated for the spectrum of the Chelyabinsk meteorite (unshocked matrix material) with increasing amounts (5% to 100%) of the shock darkened portion mixed in. The results are shown in Fig. 8b revealing a "shock darkening vector" for ordinary chondrites meteorites where the spectral change is to diminish the presence of any absorption bands. Kohout et al. (2015) explore this progression in more detail, but the purpose for our presentation is to show the orthogonal nature of the vectors for traditional "space weathering" compared with shock darkening. Most importantly for interpreting asteroid compositions is that shock darkening could alter the spectral properties so that they migrate across "line α" and thereby receive a much different taxonomic classification than their unaltered material. Simply, this adds caution to the difficulty of reliably interpreting compositional information for "featureless" asteroid spectra.

## 5. Mineralogical modeling of the NEO population

Our NEO population analysis to this point, and in previous incarnations (e.g. Binzel et al., 2004) has focused on statistics according to taxonomic class. However the breadth of the spectral data currently available (spanning the visible and near-infrared wavelengths) map out in detail the absorption bands at 1- and 2-microns for surfaces containing a measureable abundance of olivine and pyroxene. The correlations between these absorption bands and mineralogy have long been demonstrated (e.g. Cloutis et al., 1986; Gaffey et al., 1993). Having applied detailed band modeling techniques (Sunshine and Pieters, 1993) to telescopic spectra of asteroid Itokawa with specific predictive success to be an LL chondrite (Binzel et al., 2001) as confirmed by the Hayabusa sample return (Nakamura et al., 2011), we are emboldened to advance from taxonomy to meteorite analog interpretation.

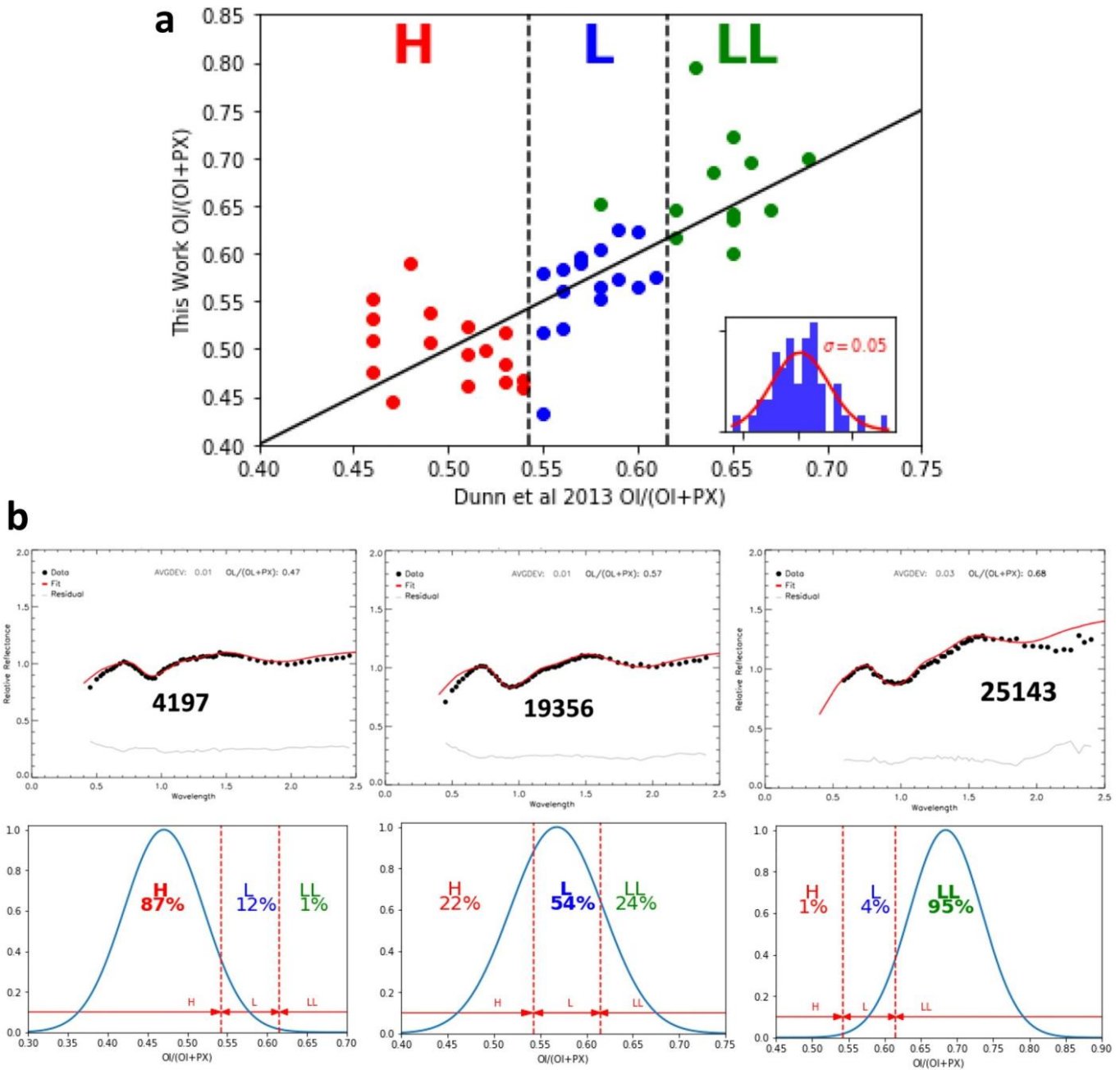

**Fig. 9.** Results for the implementation of the Shkuratov et al. (1999) model to the laboratory measurements by Dunn et al. (2010) for H, L, and LL ordinary chondrites spanning a broad range of ol / (ol+px) values. Implementation of the mineralogical model and final adjustment to the 45° line (**Panel a**) are described in Section 5.1. Boundaries values for ol / (ol+px) that discriminate objects in these three ordinary chondrite classes are shown as dashed lines: the H/L boundary is at 0.54 and the L/LL boundary is at 0.62.
**Panel b** shows the fit of our model to the spectra for three NEOs whose mineralogical solutions place them, respectively, in the H, L, and LL mineralogy ranges. Residuals of the model fit to the spectrum appear in the lower part of each spectral plot. As described in Section 5.1, we conservatively estimate that our mineralogy ratios are no better than 0.05. Thus we treat each model value for ol / (ol+px) as the mean for a probability distribution with a Gaussian standard deviation of $\sigma = 0.05$, as shown. We estimate the likelihood of a modeled asteroid being related to the indicated H, L, LL meteorite classes according to the Gaussian areas falling in each domain. Our 92% likelihood for an LL chondrite link to 25143 Itokawa is confirmed through the sample returned by the Hayabusa mission (Nakamura et al., 2011).

*5.1. Implementation of the Shukuratov model*

Within our available sample, we have 230 objects with full visible plus near-infrared wavelength coverage displaying 1- and 2-micron absorption bands and taxonomically falling within the broad range of the "S-complex." We choose to perform a mineralogic analysis across this entire suite, noting that "ordinary chondrite" mineralogies are most consistent with the subset described as S(IV) by Gaffey et al. (1993). A separate analysis of the MITHNEOS data set by Dunn et al. (2013) shows that about 10% of our sample falls well outside the S(IV) region owing to relatively higher band area ratios. (Below we detail the limitations of our assumption.) For our analysis, we implement the model of Shkuratov et al. (1999) following the code of Vernazza et al. (2008), where we seek to utilize as much of the available spectral information (shape, center, area) that can be delivered by the spectral resolution of our data spanning the full available spectral range. The

Shkuratov et al. (1999) model is a one-dimensional, radiative transfer mixing model for the spectral albedo of powdered surfaces. Light interaction with a mixed material is modeled to produce an observed reflectance spectrum. Spectra of laboratory mineralogical samples are mixed and reflectance spectra are modeled to best match an observed reflectance spectrum. Given that major features in visible and near-infrared asteroid spectra are attributed to olivine and pyroxene, these are the major components in the Shkuratov mixing model. Various physical parameters influence the observed reflectance spectra, such as effective optical path length and porosity; each of these parameters are considered in the Shkuratov modeling. For consistency in the model as we apply it here, we limited the number of free parameters available by fixing the porosity to 0.5 and limiting the effective optical path length to be greater than 50 μm. Chromite (as an opaque mineral) and clinopyroxene are found to be minor constituents in all model fits and are entered into the model at a fixed (constant) value of 0.05 while the opacity solution is bounded in the range between 1 and 10. As discussed in Section 4, space weathering is a process that alters the slope of the reflectance spectrum and therefore slope is an additional parameter that must be accounted for. To account for space weathering we use the fitting the method of Brunetto et al. (2006), where their space weathering factor "Cs" is a free parameter in our modeling solution. We implement our Shkuratov modeling by performing a broad sweep of least squares best fits to varying fractions of olivine (ol) and pyroxene (px) mineralogies, where these fractions are iterated to find the resulting model spectrum having the minimum residual fit to the input asteroid spectrum. The output product of our Shkuratov modeling is an estimate for the ol / (ol+px) ratio (and its uncertainty) that provides the best fit to the spectrum.

Most important in calibrating our implementation of the Shkuratov model has been the availability of a "gold standard" against which to test our choices of model parameters. We repeatedly applied and adjusted our model to optimally fit the meteorite sample of Dunn et al. (2010), whose direct laboratory measurement of the ol / (ol +px) ratios give "ground truth" to these meteorite spectra. Limiting the number of free parameters in the Shkuratov model, as described immediately above, created some limitations in perfectly fitting our model results to the Dunn "gold standard" across the wide range of mineralogies that our sample represents. We found that fitting the whole range of mineralogies measured by Dunn et al. (2010) and matching our model results with a line having slope 45° (Fig. 9a) thus requires a final adjustment, as follows. For the value of the ol / (ol+px) ratio delivered by our Shkuratov modeling (call it "Rm"), we achieve the best 1:1 match to the Dunn laboratory measurements by instituting a final adjustment, where our final value for the model ratio ("Rf") is given by:

Rf = 1.51Rm − 0.307

Fig. 9a plots our final values (Rf) for the ol / (ol+px) ratio compared to the Dunn et al. (2010) laboratory measurements. This final correction optimally tunes our model to the "gold standard" of Dunn et al. (2010). In doing so, we note most importantly that no correction is applied in the middle range of all our model fits, i.e. Rf − Rm = 0 for an ol / (ol+px) ratio equal to 0.602; a value that falls in the range of L chondrites. Thus our model implementation is optimized to fit the L chondrites and we can argue that our model has the greatest robustness in distinguishing whether or not any spectrum has an interpreted mineralogy falling within the range of L chondrites. A correction value becomes significant (Rf − Rm > 0.05) only for the highest ol / (ol+px) ratios (Rm > 0.70) that are typical for only the most olivine-rich LL chondrites. At the low end of ol / (ol+px) ratios, our model values (Rm) have a significant correction (|Rf − Rm| > 0.05) only for the lowest ol / (ol+px) ratios (Rm < 0.50), which falls squarely in the range of olivine-poor H chondrites. In all cases where our modeling is applied to objects having large band area ratios (high abundances of pyroxene), the resulting "interpretation" places them in the category of H chondrites. Thus a limitation of our analysis may be an over abundance of H chondrites within our sample.

The ~0.05 maximum magnitude of our |Rf − Rm| correction factor to align our final model ol / (ol+px) ratios with Dunn et al. (2010) also mirrors the residuals for the overall fit of our final model. Fig. 9a (inset) shows the distribution of our final model residuals, i.e. their scatter relative to the perfect fit to the 45° line. We find that our residuals are well fit by a normal distribution having a value of $\sigma$ = 0.05. Thus we take a conservative interpretation that each and every final value (Rf) for the ol / (ol+px) ratio is uncertain by ± 0.05. This value is also conservative relative to any phase angle effect influencing our ol / (ol +px) ratio values. For our illustrated case of 433 Eros (Section 3.1; Fig. 3), the six spectra spanning phase angles from 12 to 47° (Fig. 3a) yield ol / (ol+px) ratios within the range 0.65 ± 0.04. Fitting the phase angle corrected spectra (Fig. 3c) yields 0.65 ± 0.02; both results are comfortably within our conservative estimate of ± 0.05. Similarly, our three other test cases show minimal change for their model ol / (ol +px) ratio values. Our resulting mineralogical ratio values without [and with] phase angle corrections applied are: 1036 Ganymed 0.45 [0.42], 1627 Ivar 0.66 [0.67], 4179 Toutatis 0.61 [0.63]. These small changes, and their non-systematic direction leads us to not attempt any global correction for phase angle.

We illustrate in Fig. 9b how our conservative error for the ol / (ol +px) ratio is translated into an interpreted analog for H, L, and LL ordinary chondrite meteorites. We do not treat any individual mineralogical ratio as "an answer." Rather, each model ratio simply represents a probability distribution function having a value of $\sigma$ = 0.05. In the top row, we display three specific examples of our Shkuratov model fitting and the residuals to each fit. The three asteroids displayed 4197, 19356, and 25143 have final values (Rf) for their ol / (ol+px) ratios of 0.49, 0.58, and 0.71 respectively. The lower row of Fig. 9b shows the $\sigma$ = 0.05 uncertainty applied to each result to create a probability distribution function that spans the range of H, L, and LL ordinary chondrites meteorites. (Boundary values are given in the figure caption.) Accordingly, we conclude that asteroid 4197 has an 87% likelihood of being associated with the H chondrites, with a smaller (12%) chance of being related to the L chondrites, and yet a non-zero (1%) probability for LL chondrites. The narrow range of ol / (ol+px) ratios for L chondrites (the 'middle child' of ordinary chondrites) means that distinct characterization in to this class can be difficult. For example, asteroid 19356 has a predominant likelihood (54%) of being related to L chondrites, but has non-negligible (~20%) probabilities for the other two ordinary chondrites categories

Asteroid 25143 Itokawa is specifically chosen for illustration because its ground truth as an LL chondrite is available thanks to the success of the Hayabusa sample return (Nakamura et al., 2011). Our methodology yields a high likelihood (95%) for Itokawa being an LL chondrite, fully consistent with its pre-Hayabusa modeling and interpretation (Binzel et al., 2001). As noted in the outset to Section 5, we embrace this confirming result forged by the Hayabusa success as a foundation for confidence in exploring the source regions of ordinary chondrites meteorites (Section 7).

**6. Dynamical modeling for the NEO Population: Source regions for taxonomic classes**

Our NEO analysis up to this point has been entirely based on astronomical measurements of spectral colors, progressing from the first step of assignment into taxonomic classes to the more detailed mineralogical analysis. For the following Sections, we assert that the orbital parameters (as well as the dynamical history) for each object may be treated statistically as an *independent variable* from the spectral measurements. Thus by having spectral information that is *a priori independent* of the dynamical properties, we can statistically investigate their correlations. It must be recognized that observational selection effects can play a role in building the sample population, e.g. high albedo asteroids are more likely to be discovered by ground-based surveys than equivalent diameter low albedo objects. Thus we must clearly note that any biases (such as those introduced by diameter ranges) that are present in our observational sample will naturally be reflected in the output of the models employed. Cases where selection effects may play a determining role in the findings are addressed in our discussion of individual results.

6.1. Escape region models

In this section we explore sources for near-Earth objects, considering that they must "escape" from a region where they have otherwise held a long-term residence (presumably) traceable back to the early solar system. Here we use "source region" and "escape region" interchangeably, noting that the latter term is the correct description of the output from the models we utilize. Previous investigation of NEO sources by Binzel et al. (2004, 2015) use the five-region model of Bottke et al. (2002). Here we employ the Granvik and Brown (2018) implementation of a next generation seven-region model progressively developed by Granvik et al. (2016; 2017, 2018) that explores in finer detail the escape paths of objects from the main belt to near-Earth space, including consideration of the Yarkovsky effect (Vokrouhlický et al., 2015). As noted above, the Granvik model depicts where the object most likely departed the main-belt and became a planet crosser. The extent to which this may exactly map where the object resided for the past few Gy remains under further investigation.

While the Bottke model broadly categorized "Mars Crossers" (MC) as an escape region, the Granvik model resolves Mars crossers into contributions from the Hungaria (Hun), Phocaea (Pho), and $\nu_6$ secular resonance regions. The Hungarias are a group of high inclination ($i \approx 20°$) asteroids, isolated at the innermost edge of the main belt ($a < 2.0$ AU) by the 4:1 mean motion resonance of Jupiter. The Phocaeas are also a high inclination group ($i \approx 25°$) isolated inside of the 3:1 mean motion resonance ($a < 2.5$ AU). The $\nu_6$ secular resonance ($a \approx 2.2$ AU) defines the inner edge of the main belt across a wide range of inclinations. The Granvik model also more finely considers outer main-belt escape source regions, adding an analysis of the 2:1 mean motion resonance with Jupiter ($a \approx 3.28$ AU) to the previously considered sources from the 5:2 resonance ($a \approx 2.82$ AU) and Jupiter family comets (JFC). We also evaluated, but do not use here the source model by Greenstreet et al. (2012) that assigns probabilities to the same sources as the Bottke et al. (2002) model.

6.2. Granvik escape region model implementation

Our utilization of the Granvik et al. (2018) escape region model is illustrated in Fig. 10. Given the orbital elements and H magnitude of a single NEO as input, the Granvik model delivers as output a probability distribution function comprised by seven discrete values for each of the seven escape regions (listed above), where the sum of the discrete probabilities is equal to unity. Co-authors A. Morbidelli, M. Granvik provided the model output for each of the 1040 NEOs in our sample, noting that in the present implementation there is some overlap in the contributions between the high inclination sources (Hungaria, Phocaea) and the resonance sources. Results with regard to contributions by the high inclination sources may be tentative and are discussed accordingly below.

We illustrate our analysis methodology using the subset of 125 NEOs having a C-type taxonomy. Fig. 10a shows the mean values for the members of this subset, revealing they are dominantly delivered from the main belt to near-Earth space through the $\nu_6$ secular resonance. This $\nu_6$ dominant escape path from the main-belt into nearEarth space is well known (e.g. Bottke et al., 2002), as illustrated by the middle panel (Fig. 10b) for our full sample of more than 1000 NEOs. We

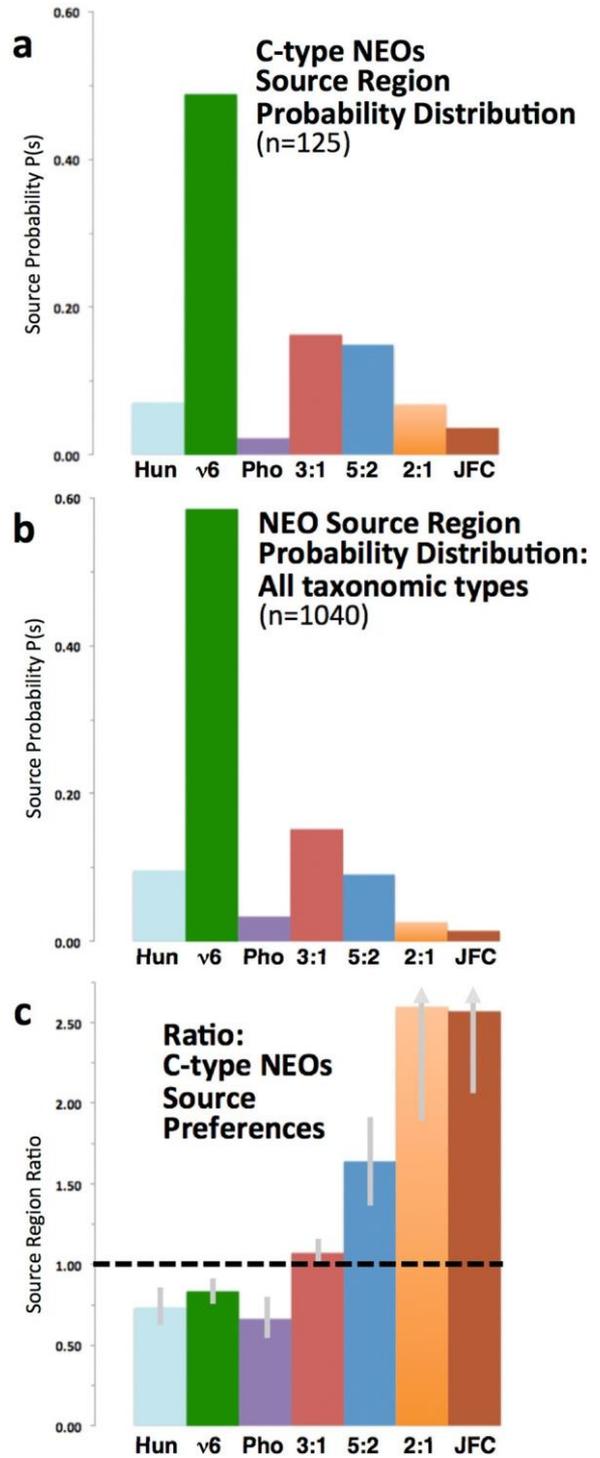

*(caption on next page)*

**Fig. 10.** Illustration of our methodology for investigating main-belt escape regions for near-Earth objects of different taxonomic types. The probability distribution function for seven different escape regions is calculated using the model of Granvik et al. (2018) as described in Section 6.1. The sum of all probabilities is equal to unity. Hun, Pho, and JFC refer to the Hungaria, Phocaea, and Jupiter family comet regions, respectively. All other regions are planetary resonances.

a) Probability distribution function for the regions where our observed subset ofC-type NEOs are delivered from the main-belt in to the inner solar system. The results plotted are the mean for 125 individual probability distribution functions.

b) Probability distribution function of escape sources for our full NEO sample.Results plotted are the mean for the 1040 individual objects in our full sample.

c) Ratio of C-type escape sources relative to the full sample (panel **a** divided by panel **b**). A ratio of unity (dashed line) indicates C-types escape from the main belt at the same percentage as our overall sample. Differences from unity by more than one standard deviation (error bars are ± 1σ) indicate regions where C-type NEOs have a relatively enhanced (or reduced) escape region preference. (Arrows indicate error bars extending beyond the bounds of the plotted vertical range.) Overall, the result shows that while C-type NEOs, like all others, escape most commonly through the $\nu_6$ resonance (shown in panel **a**), the outer belt resonances and possibly Jupiter family comet sources contribute a greater escape fraction of C-type NEOs compared to the average for the full sample.

use our full sample population, Fig. 10b, as a basis of reference for our escape region analysis. Rather than re-affirm the dominance of the $\nu_6$ secular resonance as a delivery route, our analysis seeks to ask a different question: *Do the C-type NEOs show a delivery route that is different from the average for our NEO sample as a whole?* We address this question by computing the simple ratio of the two panels (panel 10a / panel 10b), such that if C-type NEOs followed the same delivery path as the full sample, the ratio would be a constant value of unity across all source regions. To the contrary, Fig. 10c shows that C-type NEOs show a distinct non-uniform ratio (differing from unity) in their contributions from the outer belt 2:1 resonance and JFC regions.

As an estimate for whether any ratio differing from unity is statistically significant, we generate error bars for each ratio based on the standard deviation of the mean (SDOM). For the case of the subset of Ctype NEOs, the reference probabilities in Fig. 10b are treated as constants in calculating the sample mean and standard deviation (σ) from the 125 ratios for each escape region. The large sample (n > 30) allows us to invoke the central limit theorem (Kelson, 2014) for using the standard deviation of the mean (σ /√n) as an estimate of the formal error for the histogram values. We consider variations of one or more standard deviations away from unity as demonstrating some evidence for an "escape region preference" whose detailed significance is worthy of further investigation and validation; but we leave that investigation for a still larger sample reaching diameter ranges and completeness beyond the scope of the current work. 6.3. Escape region preferences for NEO taxonomic classes

Having detailed our methodology in Section 6.2, here we present and discuss our findings for the escape regions for the major taxonomic classes for NEOs. In many cases, but not all cases, this analysis is limited by the small sample size. As differing examples, our subset of ten K-type NEOs gives no interpretable result: all seven escape region bins have ratio values and large standard deviations that overlap unity. On the other hand, even though the D- and P-type NEO subsets have ten or less members, their outer belt escape region ratios are substantially greater than unity and significantly exceed estimates for their uncertainties.

Fig. 11 shows results for six classes contrasting preferential escape regions between the inner and outer main belt. As noted above, C-types show a signature of enhanced contribution that is widely distributed across the outer main-belt, consistent with the well described outer-belt provenance for C-types in the stratigraphy shown by Gradie and Tedesco (1982) and more fully described and updated by DeMeo and Carry (2013, 2014).

D-type NEOs show the highest ratio values of any type, with those values indicating strong contributions from the 2:1 resonance and Jupiter Family Comet regions. These results appear robust relative to the estimates for the error bars, but the detailed interpretation is limited by the small sample size (n = 21). P-types show one of the most localized escape region signatures standing out at the 5:2 resonance, consistent with their outer main-belt predominance. If the P-types are revealing a true signature (estimated at ~1σ), the signal is somewhat remarkable for the small sample size (n = 8). We note that each of these results is consistent with an analysis based on the Bottke et al. (2002) source model, applied to a smaller data set by Binzel et al. (2004; see Fig. 6 therein). These low albedo classes have also been investigated specifically with regard to the Tisserand parameter by Fernandez et al. (2001, 2005) and DeMeo and Binzel (2008) leading to the estimate that extinct comet nuclei likely comprise a few percent of the NEO population. This estimate of a few percent remains consistent with the recent analysis by Granvik and Brown (2018).

Inner belt signatures are seen in Fig. 11 for L-, V-, and Xe-types. We address each in turn. The L-types have been linked to bodies rich in calcium aluminum inclusions (CAIs; Sunshine et al., 2008; Devogèle et al., 2018). The particularly strong $\nu_6$ signature, even greater than the overall population (Fig. 10b) is consistent with the location of 234 Barbara (a = 2.4 AU), which these authors describe as one of their potential parent bodies. V-type asteroids and Howardite-Eucrite-Diogenite (HED) meteorites, of course, have long been linked to 4 Vesta (McCord et al., 1970; Consolmagno and Drake, 1977), confirmed through in situ measurements by Prettyman et al. (2012). The widespread distribution of small V-type asteroids across the inner asteroid belt (Binzel and Xu, 1993; Moskovitz et al., 2010) predicted and confirmed to be excavated from a large impact basin on Vesta (Thomas et al., 1997) revealed the pathway for both the meteorites and the first known V-types within the near-Earth population (Cruikshank et al., 1991). Such a widespread inner belt availability of "Vesta debris" is evidenced within our analysis showing all sources inside the 3:1 resonance giving a ratio of contributions consistent with unity (i.e. following the insertion pattern of Fig. 10b). The inner-belt dominance of V-type NEOs is further demonstrated by the statistically low ratio values calculated for the outer belt sources.

Particularly striking for a small sample size (n = 8) is the strong signature for the Hungaria region as a source for the Xe-class of NEOs. (Here we note that the Xe-class in the Bus-DeMeo taxonomy corresponds nearly identically to the traditionally known E-class first defined by Tholen (1984)). The strong link of Xe-types to the Hungaria region confirms a well-established relationship recognized by Gaffey et al. (1992) in their study of 3103 Eger, for which they give strong evidence for this object being related to the enstatite achondrite (Aubrite) meteorite class. Within the new data set presented here, (54789) 2001 MZ7 is the Xe-type having the strongest Hungaria escape region signature output from the Granvik model.

S-, Sq-, and Q-types dominate more than half of our total NEO sample and thereby most strongly determine the structure of the reference distribution shown in Fig. 10b. However the ratio to this reference distribution does not strictly follow unity. Fig. 12 reveals these classes do have distinct differences from the overall sample population: each is most strongly correlated to escaping from the inner asteroid belt rather than the outermost regions of the 2:1 resonance and Jupiter Family Comets. Interestingly, the class interpreted as having the least space weathered surfaces (Q-types, see Section 4.3) shows the greatest differences in ratio values. For these objects, the $\nu_6$ resonance has the strongest overall signature as a preferred region for being transferred from the main-belt into the inner solar system. The short timescale and efficiency of transfer from the $\nu_6$ resonance, the low inclination of their orbits and their low relative velocities for planetary encounter may be a factor allowing surface refreshing mechanisms to gain an advantage over slower but steadier space weathering processes. We address this further in Section 8.

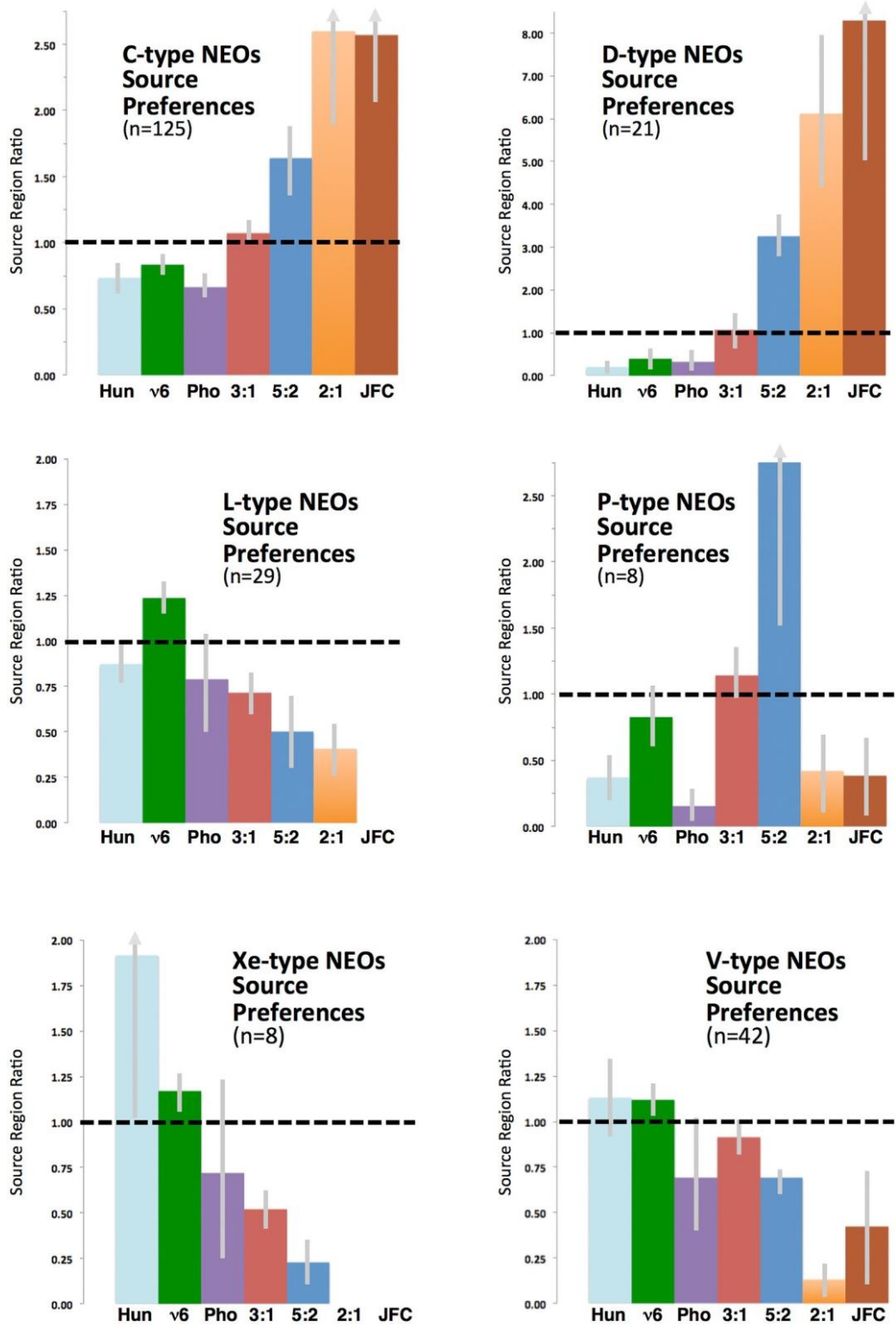

**Fig. 11.** Delivery regions for taxonomic classes of NEOs based on the escape region model of Granvik et al. (2018). Methodology and interpretations of each ratio and their uncertainties are discussed in Section 6 and Fig. 10. Clear outer main-belt escape region preferences are found for C-, D-, and P-type NEOs, with D-types strongly indicating a Jupiter Family Comet contribution. Inner main belt escape region signatures are most clearly seen for the L-, V-, and Xe-types; in each case these locations are consistent with their proposed or established specific parent bodies (e.g. Vesta for the V-types; Hungaria for the Xe-types; see Section 7).

## 7. Dynamical modeling for the NEO population: H, L, LL chondrite advance beyond taxonomy into mineralogy.

This pathway has a long **escape regions** established history (e.g. Gaffey et al., 1993) and here we follow the modeling results of Section 5.1 toward addressing the correlations and

As emphasized in Section 5, a primary motivation for the MITHN- source regions for the most commonly falling meteorites, the ordinary EOS program has been to achieve broad spectral coverage so as to chondrites. Ordinary chondrites comprise ~81% of all meteorite falls by number. Among the total fall number statistics, 43% are H-chondrites, 47% are L-chondrites, and 10% are LL-chondrites.

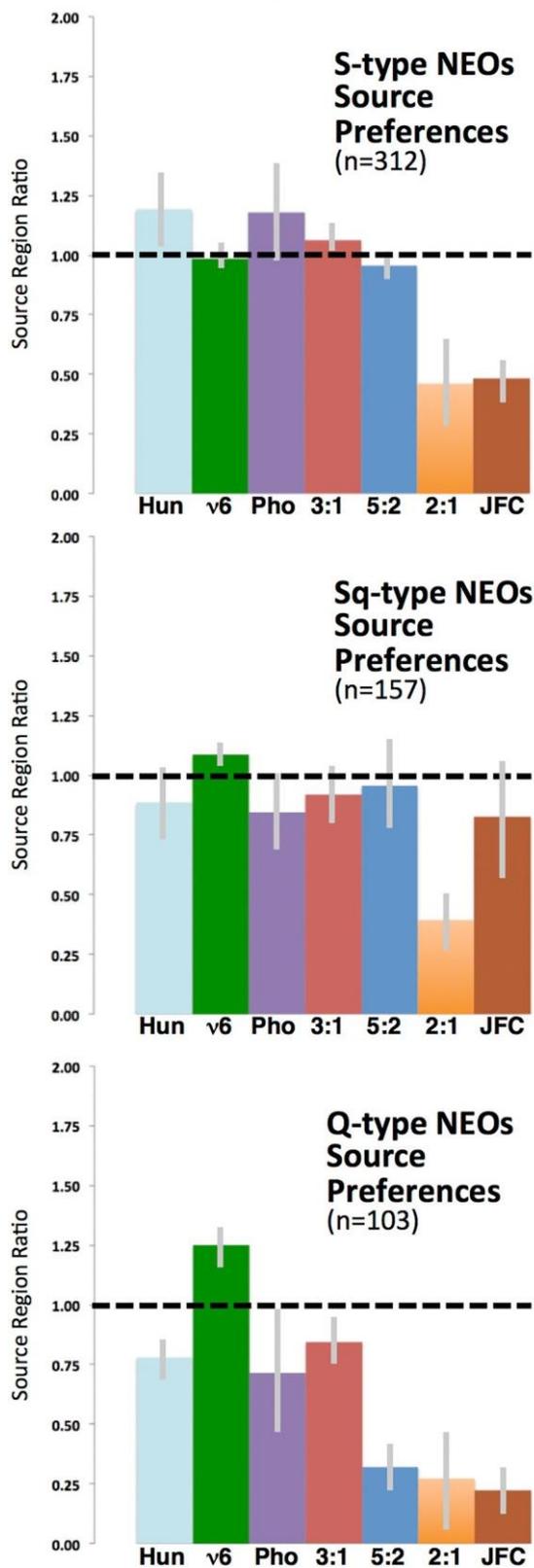

**Fig. 12.** A strong inner main-belt escape region signature is shown for the predominant taxonomic classes of NEOs, the S-, Sq-, and Q-types. Their ratio values near unity (dashed line) implies they are most abundantly delivered through the $\nu_6$ resonance, following the reference profile shown in Fig. 10b. Qtypes, interpreted as being the least space weathered objects, show an additional signature toward the $\nu_6$ resonance that may be related to their delivery into the planet crossing region with the lowest inclinations and lowest relative velocities, improving their likelihood for close planetary encounters (see Section 8).

### 7.1. Methodology for convolving probability distribution functions

For our escape region analysis, we begin by recognizing that the Shkuratov mineralogic model (Section 5.1) and the Granvik escape region model (Section 6.1) both deliver probability distribution functions. We treat each of these distribution functions as independent random variables. The Shkuratov model yields a likelihood for an Scomplex spectrum to be described as an H-, L-, or LL-chondrite while the Granvik model yields a likelihood, based on that same NEO's orbit, to have been delivered from any one of seven different source regions. Thus our method is to convolve these two probability distribution functions for revealing any possible correlation for a specific meteorite class to be preferentially derived from a specific source region.

We illustrate our method in Fig. 13. We applied the Shkuratov model to 194 S-complex NEOs and tabulated as independent variables the escape region probability distribution function for each of these objects. Logically, the sum of all escape region probability bins for the full set is 194. Dividing the accumulated sum in each of the seven probability bins by 194 yields the mean escape region probability distribution function for our sample population, with the result shown in Fig. 13b.

Hayabusa sample return asteroid 25143 Itokawa serves as our example for applying the convolution of probability distribution functions. The Shkuratov model delivers a probability of 0.95 (95%) of Itokawa being an LL-chondrite and 0.05 as an L-chondrite, consistent with Itokawa's now known LL-chondrite mineralogy (Nakamura et al., 2011). (Our modeling indicates zero likelihood of Itokawa being an Hchondrite.) Independent of its spectral properties, Itokawa's orbit properties yield from the Granvik model escape probabilities of 0.86 (86%) from the $\nu_6$ resonance, 0.05 from the Hungarias, 0.04 from the 3:1 resonance, and 0.05 elsewhere. We convolve these independent variables together, as follows: Itokawa itself carries a weight of (0.95 × 0.86=) 0.82 toward LL chondrites being delivered from the $\nu_6$ resonance. Similarly, Itokawa carries a convolved weight of (0.95 × 0.05=) 0.048 for LL chondrites being delivered through the Hungaria region and (0.95 × 0.04=) 0.038 for LL chondrites being delivered through the 3:1 resonance. Because Itokawa has a non-zero model probability of being an L-chondrite, Itokawa is also contributing statistical weight toward revealing where L chondrites come from. (For the L-chondrite case, the convolved weight is 0.043 for a $\nu_6$ source, 0.003 for a Hungaria source, and 0.002 for a 3:1 source.)

Moving forward with our example we note that of the 194 objects we model, Itokawa counts only as 0.95 of an LL-chondrite while augmenting the sample size of L-chondrites by 0.05. Each object in its own way contributes toward the three meteorite categories, with the accumulated count from the spectral modeling being 99.6 LL-chondrites, 56.7 H-chondrites, and 37.7 L-chondrites (total 194). Completing our example for "LL-chondrites" thus requires calculating the convolved weights for all 194 objects (as in the case of Itokawa, above) and summing those convolved weights in each of the seven escape region bins. We divide the total in each bin by 99.6 to yield the convolved probability distribution function shown in Fig. 13a; where the sum of the seven source probability values is equal to one.

Identifying escape region relative preferences follows the same methodology applied to taxonomic classes in Section 6.3. We use our sample of 194 spectrally modeled S-complex objects (Fig. 13b) as our reference population as we are most interested in discerning localized escape regions of H-chondrites, relative to L-chondrites, relative to LLchondrites. Our analysis of the S-, Sq-, and Q-type objects that form the basis of our ordinary chondrites spectral modeling are already shown to favor the inner belt (Fig. 12). Thus using Fig. 13b as our reference population provides a finer discernment for our meteorite sources. The resulting ratio for LL-chondrite escape regions is shown in Fig. 13c,

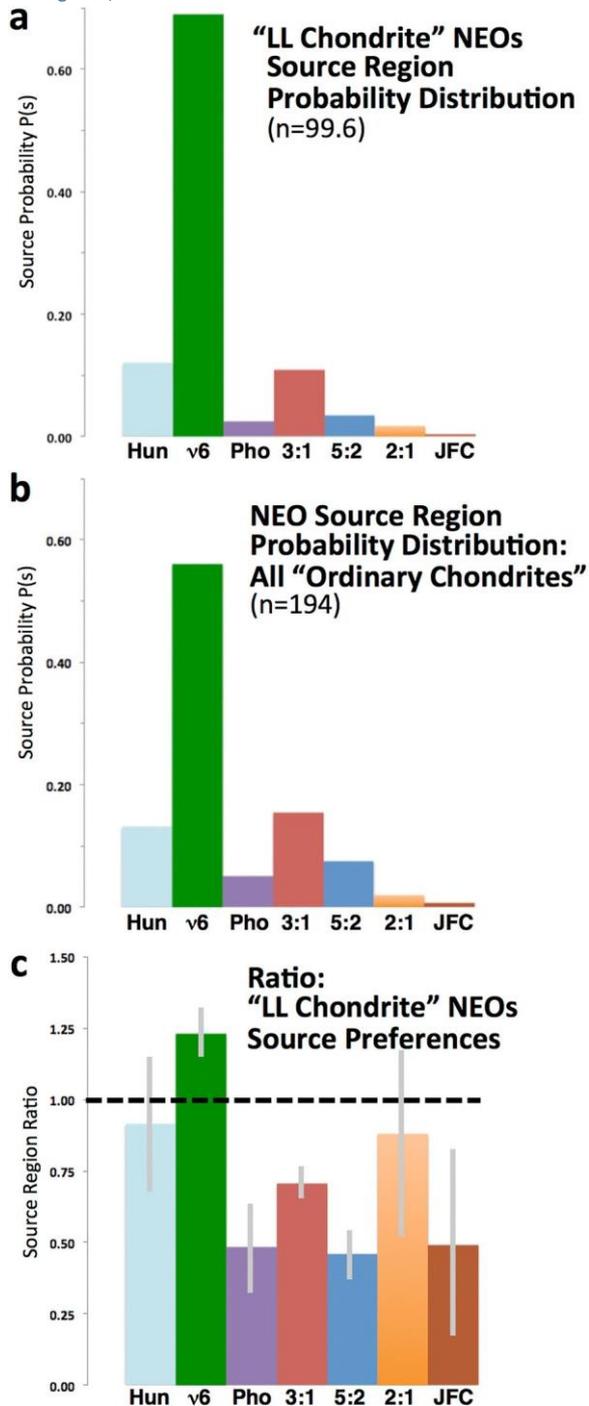

**Fig. 13.** Methodology for investigating the main-belt escape regions for near-Earth objects according to their most likely meteorite associations. We use the results of our spectral modeling (Section 5.1; Fig. 9) to deduce likely associations with "ordinary chondrites" and use "" to denote these associations are model results rather than actual meteorite samples measured in the laboratory. We compare our results with meteorite delivery results by Granvik and Brown (2018).

**a)** Escape region probability distribution function for the weighted subset (99.6 out of 194) having "LL chondrite-like" spectral properties. Escape region probability values are from the model of Granvik et al. (2018); their convolution with the spectral model probabilities is detailed in Section 7.1.

**b)** Probability distribution function of escape sources for the full sample of 194 near-Earth objects spectrally modeled as "ordinary chondrites."

**c)** Ratio of LL-chondrite escape sources relative to the full sample (panel **a** divided by panel **b**). Differences in the ratios by more than one standard deviation (error bars are +/- 1σ) relative to unity are interpreted to indicate regions where "LL chondrite-like" NEOs have a relatively enhanced (or reduced) delivery preference.

where the estimation of the 1σ error bars follows identically from the description in Section 6.2. In all cases for our ordinary chondrites escape region analysis, the largest uncertainties arise in the outermost bins for the 2:1 resonance and Jupiter family comets. As Fig. 13b shows, both of these regions have extremely low signal strength (2% and < 1%, respectively), thus making them small divisors in the analysis that follows. We display their outcomes, but do not discuss these regions further.

### 7.2. Escape regions for "H-, L-, and LL-chondrite" near-Earth objects

Fig. 14 shows a diverse set of escape region preferences for nearEarth objects that we spectrally interpret as "H-, L-, and LL-chondrites." We use " " here and in Fig. 14 to be explicitly clear that for our work we are referring to *spectral model results* for these meteorite classes, not *laboratory measured samples* themselves. Granvik and Brown (2018) present escape region results for recovered meteorite falls with welldetermined pre-entry orbits. We compare our results in Section 7.3.

For NEOs, we find the strongest and most consistent escape region signature is a preference for H chondrite-like objects being derived from the mid-regions of the asteroid belt, specifically the 3:1 and 5:2 resonances. An escape region preference for H chondrite-like NEOs at the 3:1 resonance was previously found by Thomas and Binzel (2010), where these authors used a different spectral comparison method and the source region model of Bottke et al., (2002). Thus we consider a mid-belt preferential escape source for the H-chondrites to be a robust result. Our results are consistent with the Gaffey and Gilbert (1998) proposition of asteroid 6 Hebe as the H-chondrite parent body. However, Hebe itself may not be uniquely required as Vernazza et al., (2014) show an abundance of H chondrite-like bodies in the vicinity of the 3:1 resonance that provide multiple possibilities as H-chondrite sources. Marsset et al. (2017) further argue that adaptive optics imaging of Hebe's physical shape does not reveal a significant excavation feature as might be expected for a major meteorite source body (for example, Vesta: Thomas et al., 1997). The presence of an escape region signal for the 5:2 resonance within our "H-chondrite" group adds a more distant component that supports a more broadly distributed source (multiple bodies) for H-chondrite meteorites. Similarly, our "Hchondrite" group is the only one of the three to show a preferential signature for the high inclination Phocaea region; we address this in Section 7.3.

For our "L-chondrite" NEO sample (Fig. 14; middle panel), we reiterate that an escape region ratio equal to one implies that the delivery model most closely aligns with the reference population shown in Fig. 13b. Thus the ratios near unity for the "L-chondrites" show that this group shows no particular deviation from the norm for our total sample, where the $v_6$ resonance predominates with about 60% of the escape region signal, followed by the Hungaria zone and 3:1 resonance with about 15% each. Binzel et al. (2016) reported a preliminary analysis using the Bottke et al. (2002) source region model that produced an "Lchondrite" escape region preference signal for the 5:2 resonance that exceeded the estimated uncertainty by more than 1σ. However our full sample analysis with the Granvik model offers a 5:2 escape region preference that is possibly significant only at the 1σ-level. This outer belt region, and in particular the Gefion family adjacent to the 5:2 resonance have been proposed as a supply source for the L-chondrite meteorites

(Nesvorný et al., 2009), with a recent analysis by McGraw et al. (2018) suggesting the Gefion region may be more compositionally diverse. Our analysis is not inconsistent with the 5:2 resonance being a preferred L-chondrite delivery source, but does not offer a strong conclusion.

"LL-chondrites" within our NEO sample population show the most consistent escape region signature, with a clear dominance for delivery through the $v_6$ resonance. Noting that the reference population (Fig. 13b) begins with a nearly 60% delivery factor through the $v_6$, our results (Fig. 14, lower panel) show that LL chondrite-like NEOs are another 1.2 times more likely to have the $v_6$ resonance as their escape source. That dominance by the $v_6$ is further shown by the Phocaea, 3:1, and 5:2 resonance regions all falling well below their nominal contributions. We consider the $v_6$ escape preference for "LL-chondrites" to be a robust and secure result, confirming findings by Vernazza et al., (2008), de León et al. (2010), and Dunn et al. (2013) who also showed the Flora family asteroids adjacent to the $v_6$ resonance are fully consistent with LL meteorite mineralogies. The substantial fraction of LL chondrite-like bodies in the NEO spectral sample (effectively 100 of the

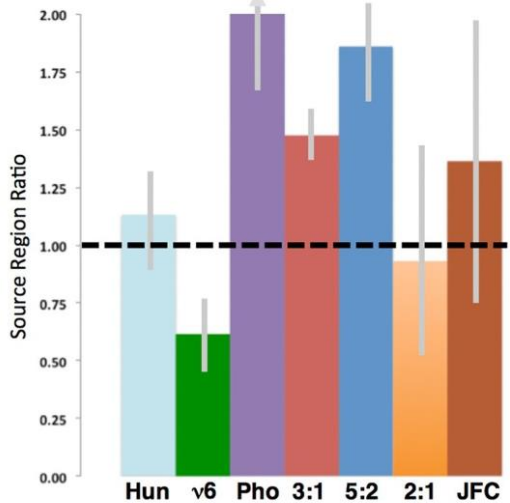
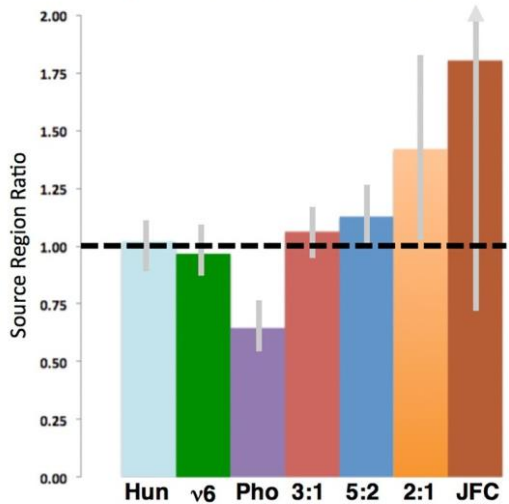
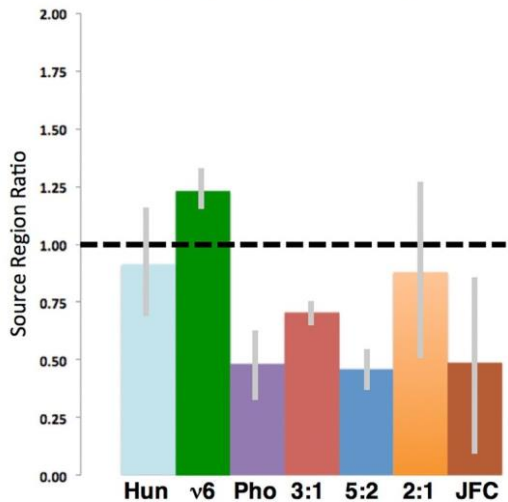

**Fig. 14.** Distinctly different escape region delivery preferences are displayed for near-Earth objects whose spectral properties are most analogous to "H-, L-, and LL-chondrites." (Methodology is the same as Fig. 13. Quotes denote these are compositional interpretations, not actual meteorite samples.) Objects most like H-chondrites have their strongest delivery signature from the mid-asteroid belt, including a transition to the inner solar system through the high inclination Phocaea region. (We note that this Phocaea signature may be a byproduct of the strong 3:1 resonance contribution. L chondrite-like objects give their strongest values toward the outer-belt, but none are significant with respect to the estimated errors in the method. LL chondrite-like objects show a strong signature toward favoring the inner-belt as a delivery source region, most notably through the $v_6$ resonance. Section 7.2 describes the correlation of these results with previous work and with direct analysis of meteorite orbits.

194 objects, or about one-half) contrasts with LL-chondrite contributing only 10% of all ordinary chondrites falls (8% of all meteorite falls). This discordance between LL abundance as NEOs and LL paucity as meteorite falls remains an unsolved problem in reconciling asteroid-meteorite connections described as the "*Asteroids IV* scenario" by Binzel et al. (2015). A review of argon-argon impact ages for ordinary chondrites (Swindle et al., 2013) also gives insight, noting that the LLchondrite meteorites show the oldest impact ages, peaking most strongly at an age > 4000 Ma. In contrast H-chondrite meteorites show a broad distribution of peak ages < 1000 Ma, with L-chondrites showing the youngest peak at < 500 Ma. Thus the timing for the most recent production of an abundance of small fragments (of order meters in size) may determine the proportions of H-, L-, and LL-chondrite meteorite falls; the timing of the argon-argon ages supports the order L (47%), H (43%), LL (10%) from most common to least common falls. It may simply be our NEO sampling is driven by less democratic transfer processes than meteorite sampling. For example compared to NEOs, the meter-sized bodies dominating the meteorite flux can have Yarkovsky drift path lengths that are a significantly larger fraction of the

(e.g. Moskovitz et al., 2017b) can investigate the validity of this hypothesis. If correct, the smallest (meter-sized) NEOs in Earth crossing orbits that most often deliver meteorite samples should be less dominated by LL chondrite mineralogies and have H-, L-, or LL-interpreted mineralogies that more closely match the proportions of meteorite fall statistics.

Table 2
*7.3. Escape region comparison for near-Earth objects and meteorites*

We compare our escape region results for NEOs with results for recovered meteorite falls analyzed by Granvik and Brown (2018), where they apply the source region model from Granvik et al. (2016). The complementarity is rather complete: for NEOs the orbits are precise but their compositions are interpretations. For recovered meteorite falls, the orbits must be interpreted from meteor camera data but the compositions are precise. For NEOs, more than 1000 objects have some interpretable compositional information (taxonomy), while for ~200 we have sufficient spectral information to

Associations of stony meteorite classes and asteroid types used as input data to create a schematic for meteorite source regions (Fig. 15). Our methodology and results for preferred source region analysis is described in Sections 6 and 7. References related to the inferred associations are given in the last column, but for reasons of space limitations, this reference list is certainly incomplete. Percentages are in terms of the total mass of recorded falls for all stony meteorite so that the sum total approaches 100%. Note that iron, stony-iron, and other metal-rich classes account for a comparable total mass of meteorite falls, but these classes are not included in the accounting here nor are they depicted in the cumulative curves of Fig. 15 (See References: Gaffey et al. (1992, 1993), Binzel et al. (1996), Gaffey and Gilbert (1998), Thomas et al. (2010), Vernazza et al. (2008, 2014, 2015), Nesvorný et al. (2009), Fornasier et al. (2015), McGraw et al. (2018), Vokrouhlický et al. (2017), Granvik and Brown (2018), Bowell et al. (1978), Vilas and Smith (1985), Love and Brownlee (1993), Cuk et al. (2014), McCord et al. (1970), Consolmagno and Drake (1977), Cruikshank et al. (1991), Binzel and Xu (1993), Prettyman et al. (2012), Sunshine et al. (2008), Devogèle et al. (2018), Bogard and Johnson (1983), Smith et al. (1984)

| Stony Meteorite falls by class (% of falls, by mass) | Asteroid type association(s) | Preferred source region(s) or source body | Meteorite-asteroid association references (incomplete listing) |
|---|---|---|---|
| H chondrites (42%) | S-, Sq-, Q-types | 3:1 resonance | Gaffey et al. (1993), Binzel et al. ((1996), Gaffey and Gilbert (1998); Thomas et al. (2010), Vernazza et al. (2014); This work, Fig. 14. |
| L chondrites (26%) | | 3:1 resonance, possibly out to 5:2 resonance. Gefion family? | Nesvorný et al. (2009), Fornasier et al. (2015), McGraw et al. (2018); This work, Fig. 14. |
| LL chondrites (10%) | | $\nu_6$ resonance, Flora region | Vernazza et al. (2008), Vokrouhlický et al. (2017), Granvik and Brown (2018); This work, Fig. 14. |
| Carbonaceous chondrites and IDPs (10%) | C-, D-, P-complex | Outer belt | Bowell et al. (1978), Vilas and Smith (1985), Love and Brownlee (1993), Vernazza et al. (2015). |
| Aubrites (5%) | E-, Xe-types | Hungaria region | Gaffey et al. (1992), Cuk et al. (2014). |
| Howardite, Eucrite, Diogenite (main-group oxygen isotope HEDs; 3%) | V-type | Vesta | McCord et al. (1970), Consolmagno and Drake (1977), Cruikshank et al. (1991), Binzel and Xu (1993), Prettyman et al. (2012). |
| Rich in Calcium-aluminum inclusions (CAI-rich; ? %) | L-type | Inner main-belt, Barbara family | Sunshine et al. (2008), Devogèle et al. (2018). |
| Shergottites, Nakhlites, Chassignites (SNC; 0.2%) | Mars Trojan Sa-types? | Mars | Bogard and Johnson (1983), Smith et al. (1984), Polishook et al. (2017). |

Polishook et al. (2017)).

asteroid belt's overall dimension. Thus meteorite-sized parent bodies, as compared to NEOs, may be sampled somewhat broadly (much more democratically) from across the asteroid belt. In contrast, for increasingly larger-sized objects (our telescopic NEO sample is mostly ~100 m objects and larger) Yarkovsky drift path lengths are generally shorter (or at least drift more slowly). Thus for the sizes of objects in our NEO sample, delivery efficiency into resonance escape regions may have an important dependence on heliocentric distance. For a large (> 100 m) object, proximity to a resonance and a maximum Yarkovsky drift rate (closer to Sun) may significantly bias the observable sample. In particular, the abundance of LL chondrites bodies in the Flora region, and the efficiency of the $\nu_6$ resonance can result in "LL chondrite" NEOs being the most prevalent among NEO-sized objects, even if LL-chondrite meteorites are the least common of the three groups to fall. Spectroscopic surveys driven toward sampling the smallest available NEOs, even below 10 m interpret specific meteorite sub-classes (Section 7.2). For meteorites, the full statistics are available for far fewer objects (~25). Borovička et al. (2015) provide a recent summary of the available meteorite fall data and (Dumitru et al., 2017) summarize possible meteor shower associations.

We find that these complementary studies produce consistent interpretations and insights toward bridging the asteroid-meteorite divide. Like the NEOs, Granvik and Brown show meteorites have a dominant source of entry into near-Earth space through the $\nu_6$ resonance. One unresolved difference is a somewhat higher relative contribution of the Hungaria region toward meteorite delivery as compared to the NEOs; this may be related to the close coupling of the high inclination source to the nearby resonance. Granvik and Brown find that two (of the three) carbonaceous chondrites meteorites with available orbits have their greatest escape region weights toward the 3:1 resonance or beyond, consistent with our findings for C-type

NEOs. Similar to our NEO findings for "H-chondrites" having a mid-belt delivery preference, Granvik and Brown show a significant weighting for 12 H-chondrite meteorites toward the 3:1 resonance. Three of these meteorites, Mason Gully, Pribram, and Benesov have 3:1 resonance escape region probabilities in the range of 60 to 80 percent. For the six available L-chondrite meteorite falls, their escape regions are most strongly weighted from the 3:1 resonance inward, consistent with "Lchondrite" NEOs that follow closely the reference delivery profile displayed in Fig. 13b. Only one LL-chondrite meteorite fall has sufficient measurements to estimate its pre-entry orbit. That object (rather famously being Chelyabinsk) has an 80% likelihood of delivery from the $v_6$ resonance, consistent with the seemingly well-sampled population of "LL-chondrite" NEOs.

*7.4. Schematic mapping of stony meteorite source regions*

With the mapping of ordinary chondrite meteorite source regions presented here, as well as the recent results by Granvik and Brown (2018), and decades of contributions by numerous researchers (see summary in Table 2), we present in Fig. 15 a "big picture" schematic tracing stony meteorite types to their source regions. We limit ourselves to stony meteorites owing to the ambiguity of visible and near-infrared spectral interpretation (having featureless spectra, and a range of albedos) for iron and stony-iron asteroids. Similarly, we limit our interpretations of achondrite meteorites to two well-established cases (Mars and Vesta). We do not consider this schematic definitive as we recognize that the history of planetessimal scattering in the early solar system (e.g. Gomes et al., 2005; Morbidelli et al., 2007; Walsh et al., 2011) makes the distribution of sources much more complex than can be shown. (See for example the complexity of the distribution of taxonomic types illustrated in Fig. 3 of DeMeo and Carry 2014.) We additionally, and most importantly emphasize that our schematic is based on near-Earth objects that are in a different size range than meteoroids; this may account for a difference in source region conclusions when considering the morning or afternoon time of fall (Wisdom, 2017). As we note in Section 7.2, meter-sized meteorite source bodies can have substantially greater drift path lengths compared with NEOs.

The pie chart (top left; Fig. 15) shows our use of the Granvik et al. (2018) model, where the larger sizes of NEOs (compared with meteorite source bodies) is a factor in our depicting how the $v_6$ resonance dominates the delivery for all sources. What we attempt to illustrate in our schematic is the manner in which source regions show a preference in excess of the averages, as detailed in discussions of our methodology in Sections 6 and 7 of this paper.

For Fig. 15 we follow the illustration method of DeMeo and Carry (2014) where we base our meteorite source distribution on *mass* rather than by number of falls. We tabulate our mass estimates using the *Meteoritical Bulletin* database https://www.lpi.usra.edu/meteor/metbull.php. In addition, we also consider the mass of carbonaceous material contributed by interplanetary dust particles (IDPs) following the estimates by Love and Brownlee (1993). Table 2 tabulates an overview summary the asteroid-meteorite associations and source region estimates from the literature that influence our depiction.

**8. Dynamical modeling for the NEO population: Surface refreshing processes**

*8.1. Definition of a space weathering parameter*

For this final section, we quantitatively investigate the interplay between competing processes of surface weathering versus surface refreshing. For this investigation, we take further advantage of correlating the independent variables of spectral measurements and dynamical properties. Here we make specific use of the space weathering vector within the principal component space of the Bus-DeMeo taxonomy as described in Section 4.3 and presented in Fig. 8. We define our "Space Weathering Parameter" as the scalar magnitude of this vector measured from "line η", noting that Δη = 0 places an object exactly on the formal border between the Sq- and Q-type classes:

$$\text{Space Weathering Parameter} = \Delta\eta = \frac{-\frac{1}{3}PC2' + PC1' + .50}{1.0541}$$

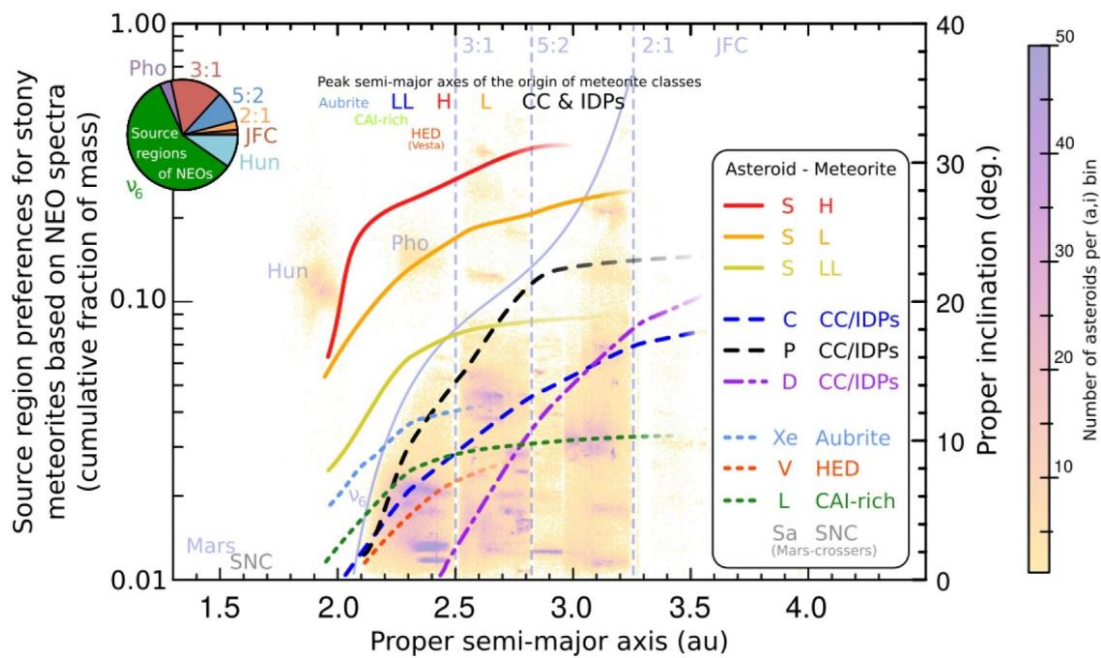

**Fig. 15.** Schematic representation for "Where do meteorites come from?" based on tracing the preferential source region signatures (main-belt escape routes) of nearEarth objects. We correlate these escape routes with the inferred meteorite associations based on our spectral measurements. The inferred meteorite associations are shown in the key at right, with more detailed references compiled in Table 2. This interpretive diagram depicts only stony meteorites and is scaled by the total mass of all stony meteorite falls. (Iron meteorites, which do not reveal diagnostic spectral features in the wavelength range of our NEO telescopic data, are neither depicted here nor included in calculating the total mass we consider.)
For each stony meteorite class, the displayed curve is cumulative, growing larger with increasing semi-major axis, and leveling off at their maximum value. (For example, H chondrites comprise about 40% of the total mass of stony meteorites, L chondrites 30%, and LL chondrites 10%.) The steepest rise in each curve is diagnostic for their strongest escape route signature from the main-belt into near-Earth space based on the model of Granvik et al. (2018), as described in Section 6.1. Labels across the top indicate the semi-major axis order for the preferred escape route locations of the major classes of stony meteorites. Most important to note is that these escape preferences are superimposed on the overall delivery distribution of the Granvik et al. (2018) model, as shown in the pie chart upper left. The escape route (source region preferences) displayed by the cumulative curves are deduced using the methodology described in Section 6.3 and illustrated in Fig. 10. (Figure design credit: B. Carry).

Our "Space Weathering Parameter" is identical to what is termed as the "spectral parameter" used by Binzel et al. (2010; see their Supplemental Information). The distance Δη is measured along the dimension parallel to the slope of line α, owing to their correlation discussed in Section 4.3.

In the following section we correlate the spectral properties of Q-, Sq-, and S-type NEOs relative to perihelion distance. We note that spectral slope, as a proxy for space weathering, has been previously correlated with perihelion distance by numerous researchers, including Marchi et al. (2006a,b). While our findings (Section 8.2) are consistent with these earlier studies, our Space Weathering Parameter is completely independent of spectral slope and instead tracks alteration in spectral features (using the full 0.45–2.45 μm spectral range) that mimic laboratory measured space weathering alteration effects (Brunetto et al., 2006). (Section 3.3.2 discusses in detail the independence from spectral slope of the PC1' and PC2' principal components.) Thus we present an independent analysis of space weathering trends that confirms earlier findings. Here we seek to extend the limits in interpreting these results.

*8.2. Space weathering and resurfacing processes in the inner solar system*

Fig. 16a presents our raw data for 195 NEOs falling in the category of Q-, Sq-, to S-types that are thought to follow the progression from the freshest (presumably resurfaced) objects to those having the most weathered surfaces (Binzel et al., 1996). Near-zero and negative values of the Space Weathering Parameter generally carry the Q-type taxonomic label. Those having the largest values for the Space Weathering Parameter fall into the category of S-types. Intermediate between these two are the objects typically classified as Sq-types. As a first step in our analysis, we abandon these familiar taxonomic labels and use the Space Weathering Parameter (Section 8.1) as a continuous variable to bin the data along the dimension of increasing perihelion distance. These binned results show a clear progression toward increasing space weathering with increasing perihelion distance, a result (as noted above) shown previously by Marchi et al. (2006a,b) based on spectral slope.

As a basis for discussion, we assume a simplistic scenario where the surfaces of NEOs in the inner solar system are caught in a "tug-of-war" between the unceasing process of space weathering versus episodic resurfacing events. Fig. 16b depicts arbitrary boundaries for the purpose of facilitating our discussion, placing the data into three regimes: objects that have undergone some (presumably recent) process to "refresh" their surfaces, objects that have extensively weathered surfaces possibly to the point of saturation, and objects in an intermediate state between these two. We consider each of these regimes in turn under the supposition that the degree of space weathering (as estimated by our parameter) is a function of surface exposure time. Certainly this is an over simplification as susceptibility to space weathering is likely subject to other factors such as compositional variations and the abundance (or absence) of regolith itself.

For the region labeled "Saturated Space Weathering" (Fig. 16b), we consider that surfaces could reach this state through different histories. Most straightforwardly, these could be surfaces that have rested passively with regolith grains exposed to the space environment for an extended length of time, likely ≥ $10^6$ years. (See Brunetto et al., 2015 for a discussion of space weathering timescales.) It is also possible that surfaces with the greatest values for the Space Weathering Parameter may have become "saturated" by an active process involving repeated resurfacing events followed by extended periods of weathering. Under this scenario, each new orientation exposing a regolith grain to space gives the opportunity for that exposed facet to become weathered before being rearranged again. After multiple re-arrangement events, surface grains become weathered on all sides (the culinary equivalent of being "sautéed") and thus are saturated by space weathering. Once saturated, further re-arranging of the regolith results in no new exposure of any "fresh" grains. As a consequence, an NEO with a "saturated" regolith could undergo one or more additional resurfacing events that afterwards yields no discernable change in surface spectral characteristics. In other words for a saturated surface, a subsequent major resurfacing event may not

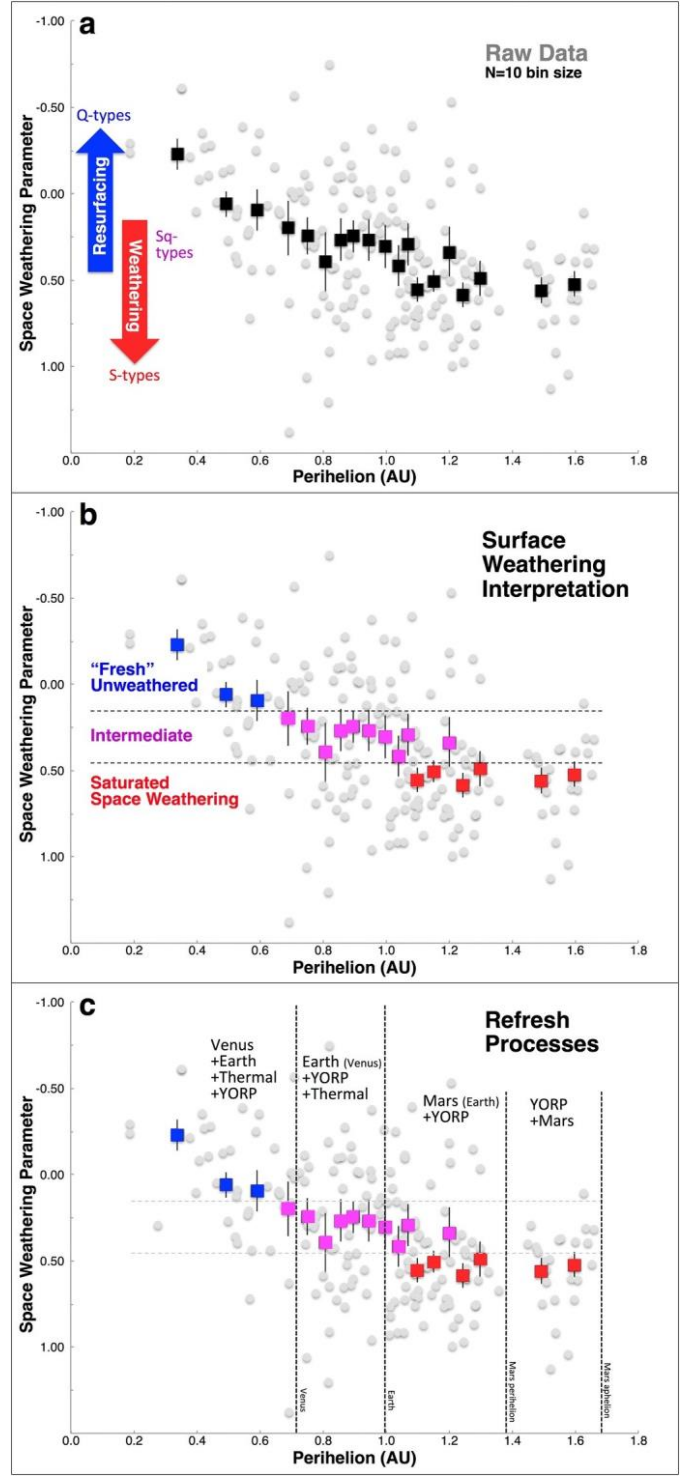

*(caption on next page)*

**Fig. 16.** Within the inner solar system, multiple resurfacing processes appear to operate in a direction opposite to that of space weathering. We define a "Space Weathering Parameter" (Section 8.1) that measures the alteration of spectral bands (over the full wavelength range 0.45–2.45 μm) and that is independent of spectral slope. **(a)** Spectral

characterization of space weathering for 195 nearEarth and Mars-crossing objects plotted as a function of their perihelion distance. (Distances plotted are for current orbits; orbital variations are discussed in the text.) Nominally these objects fall into S-, Sq-, and Q-type taxonomic classes. To explore possible trends, we bin the data in groups of 10 in the order of increasing perihelion distance. Error bars for each bin are the ± 1σ value for the standard deviation of the mean. **(b)** An interpretation is offered for different regimes depicting the interplay between space weathering and resurfacing. The regime of (presumably recent) refreshed surfaces appears at the top, while the regime for highly weathered surfaces, possibly to the point that regolith grains are weathered on all facets ("saturated"), appears at bottom. In the intermediate regime, surfaces may be in states of progressively increasing weathering with time or be representative of a variegated mixture (for example, asteroid Itokawa, Fig. 17). **(c)** An interpretation for the dominant resurfacing processes that may be operating in different orbital zones delineated by perihelion distances. Within each zone, labels from top to bottom list the hypothesized efficiency of available processes, from greatest to lesser, intended as a framework for future work. (See Section 8.2, for detailed process descriptions and associated references.) Examining the sample distribution from larger to smaller values of perihelion distance appears to show some step-wise changes in the frequency of occurrence for fresh surfaces, possibly indicative of onset boundaries for planetary encounter tidal effects. The infrequency of weathered surfaces for perihelion distances inside of Venus is particularly pronounced, possibly owing to the efficiency of refreshing processes dependent on solar flux and the particularly short survival lifetimes for these objects.

---

produce a new surface that our measurements (interpreted by our space weathering parameter) would categorize as being "fresh."

Under our discussion scenario, the objects categorized as "intermediate" may be reside in the most interesting regime – being actively caught in the "tug-of-war" between continuous weathering exposure and episodic resurfacing events. Nominally one might consider that intermediate objects are in a progressive state of weathering that is dependent only the time elapsed since the most recent resurfacing event. However the situation for how an object arrives in this "intermediate" category is most likely much more complex, as supported by direct evidence from the Hayabusa mission (Saito et al., 2006). As shown in Fig. 17, the surface of asteroid Itokawa is revealed to be variegated such that while most of the surface appears thoroughly weathered (dark regions), some bright "fresh" patches dot the landscape. Itokawa's surface demonstrates that resurfacing does not have to be a global uniform (100% surface area) process, but can occur on local scales. We consider these below.

Fig. 16c illustrates the multiplex of processes that may be involved in refreshing asteroid surfaces on both local and global scales. We evaluate these processes below and offer as a hypothesis their relative efficiencies in each of the dynamical zones depicted. We consider that deducing their actual efficiencies as a function of intrinsic physical properties (e.g. size, shape, regolith particle size distribution) and dynamical location remains an open question. The hypotheses we offer here are intended as a framework for spurring ongoing investigations.

YORP spin-up may be the resurfacing process that is operative over the greatest range of heliocentric distances. As reviewed by Vokrouhlický et al. (2015), re-radiation of accumulated solar flux can act to torque the rotation of a small body to the point of reaching the maximum spin rate at which regolith particles can no longer be retained by self-gravity or cohesion against centrifugal forces. Under this scenario, an object could be partially refreshed if there is simply pole-toequator migration of grains (maximum rotational velocity is at the equator), or a surface could be totally refreshed if spin-up reaches the point where most or all regolith is shed. At the extreme limit of spin-up, an object may fission and separate as two bodies forming asteroid pairs. Indeed, Polishook et al. (2014a) measured a Q-type/fresh spectra on a few members of asteroid pairs located in the main belt, supporting this scenario. At the extreme limit of spin-up, an object may fission and separate as two bodies forming asteroid pairs. At present, the manner in which dust may escape, resettle, or be retained electrostatically is uncertain (Polishook et al., 2014b). Graves et al. (2018) model the sizedependence and effectiveness of YORP spin-up at 2.2 AU, in the inner main-belt beyond the orbital distance of Mars. Based on the effectiveness shown by these authors, we concur that YORP spin-up is also an important contributor to surface refreshing processes, which we evaluate throughout the dynamical regions depicted in Fig. 16.

Boundary limits shown for the orbits of the terrestrial planets in Fig. 16c may illuminate the role that tidal resurfacing plays as a consequence of close planetary encounters (Nesvorný et al. 2005, 2010; Binzel et al., 2010). It is important to note these boundaries are not exact: our analysis uses current osculating orbital elements for calculating the perihelion distances shown. Secular and periodic oscillations in orbital elements, as well as planetary encounters, can cause NEOs to phase in and out of being planet crossers. At the greatest perihelion distances, our observed sample of 195 NEOs reveals very few "fresh" surfaces among objects having orbits crossing only the aphelion distance of Mars. At that distance it may be the case that the slow-andsteady (YORP) process wins the "race" as the more efficient process relative to infrequent encounters with Mars. (Note that we do not take into account any consideration of resurfacing by impact cratering or mutual asteroid-asteroid collision events.)

As measured by the Space Weathering Parameter, our NEO sample does show a step function increase in the number of "fresh" surfaces at the transition across the perihelion distance of Mars (Fig. 16c). While YORP-driven resurfacing operates as a continuous process (increasing efficiency with decreasing heliocentric distance), planetary encounters operate more nearly as a step function as may be evidenced in the data. Only about one-half of our sample of "fresh" objects in the zone between Earth and Mars reveal (through long-term integrations) orbital intersections with Earth. Thus as found by DeMeo et al. (2014), if planetary encounters are the dominant resurfacing process, Mars encounters must also be effective. Overall for this zone, we hypothesize that Mars-EarthYORP is the order of process efficiency where the step function nature of the transition is our basis for conjecturing that planetary encounters jump ahead of YORP in relative efficacy. Since all objects in this region cross the orbit of Mars, but not all intersect with Earth, we conjecture Mars as the more important planetary protagonist in this zone.

A similar step-like increase in the occurrence of "fresh" surfaces may be present (Fig. 16c) for our sample as viewed moving inward across the boundary of Earth's perihelion distance. Within this Earth-Venus zone we view this step-up as indicating a likely dominant role for Earth encounters as a resurfacing process (Binzel et al., 2010). Venus may play a role for objects plotted just beyond its perihelion limit (where orbits can phase in and out of being Venus crossers). Even though many of the NEOs in the Earth-Venus perihelion zone also remain in orbits that cross Mars, we discount the role of Mars in this zone owing to its smaller mass; we fully expect encounters with the larger terrestrial planets to dominate in this region.

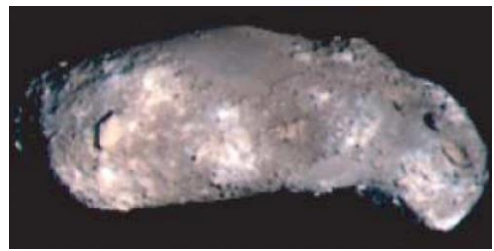

**Fig. 17.** High contrast image of near-Earth object 25143 Itokawa obtained by the Hayabusa mission (figure reproduced from Saito et al. 2016). Itokawa shows variegations over a wide range of space weathering. In this color composite, the most weathered surfaces appear dark while the freshest regions are bright. The fresh surfaces may be exposed by downslope movement or by diurnal thermal cycles breaking down the regolith and exposing fresh grains (Delbo et al., 2014). From an Earth-based observing perspective, the globally averaged reflectance spectrum (Sq-type) indicates Itokawa is in an "intermediate" regime for space weathering.

As we evaluate objects having perihelion distances inside of 1 AU, we add consideration of the role of diurnal thermal cycling. Delbo et al. (2014) demonstrate that day-night temperature fluctuations create thermal stress fractures in surface materials, cracking open the outer layers of previously solid regolith constituents (ranging from pebbles to boulders) and exposing new fresh material. Because this process is most effective where high daytime surface temperatures can be reached, thermal cycling is likely most effective for objects that reach the lowest perihelion distances. Thermal cycling is a local process, exposing fresh surface material at scales of cm (pebbles) to meters (boulders) and may account for much of the variegation seen across the surface of asteroid Itokawa (Fig. 17). We justify our arbitrary choice of adding consideration of thermal cycling as an important process inside of Earth's perihelion distance based on a feature present in our sample data: the Space Weathering Parameter shows its maximum dispersion in the Earth-Venus zone (Fig. 16c). This region may be the most active "tug-of-war" zone where surface alteration events are frequent enough to show the range between the "freshest" surfaces and those that have been repeatedly shaken to the point of complete space weathering saturation (effectively, fully sautéed grains).

Both YORP thermal spin-up and diurnal thermal cycling have an increasing efficiency with decreasing heliocentric distance, as may be evidenced by the distribution of our NEO sample having perihelion distances inside the orbit of Venus. In Fig. 16c, looking inward across the Venus perihelion boundary we note a stepwise drop in the occurrence of highly space weathered surfaces. We consider the lack of weathered surfaces (and higher abundance of fresh surfaces) inside the perihelion distance of Venus to be the culmination of multiple factors: planetary encounters are at their most frequent rate of occurrence and processes driven by solar flux are at their greatest effectiveness. A dominating consideration may also be a decrease in survival lifetime for objects at low perihelion distances, where their extermination is driven by factors including: frequent planetary encounters that remove them by planetary accretion (impacts) or gravitational ejection, YORP spinup or thermal cycling to the point of complete dissipation, and orbital eccentricity evolution that plunges them into the Sun (Wisdom 1983; Farinella et al., 1994). Granvik et al. (2016) also found a paucity of NEO orbits at small perihelion distances, further highlighting the reality of these extermination processes. Thus for objects penetrating inside the orbit of Venus, where resurfacing processes are at their maximum efficacy and survival lifetimes are at their shortest, we may fully expect to see a paucity of long-exposed and saturated weathered surfaces.

## 9. Conclusions and future work

With this work, we hope to show the value of a large data set compiled with a consistent set of observational procedures and delivering detailed spectral coverage over a wide wavelength range. Such a data set allows the investigation of important evolutionary trends, such as space weathering and shock darkening, whose full understanding

**Supplementary materials**

remains to be achieved. In all cases, we emphasize the importance of moving beyond the initial stage of taxonomic classification in to quantitative diagnostic parameters either for demonstrating trends or for detailed mineralogical analysis. The latter forges the link between telescopic remote sensing of objects in space with the measurement precision achievable for meteorites in Earth laboratories. While some new insights in linking meteorites to their delivery locations may be illuminated here, exploring the interplay between spectral properties and dynamical histories remains an open opportunity. Main-belt processes of collisions (e.g. family formation) likely play a very strong role in supplying the flux of near-Earth objects, perhaps in a highly timedependent way. Our analysis here provides a snap shot of the current inner solar system object population whose exploration, understanding, and possible utilization continues to beckon.

**Acknowledgments**

Part of the data utilized in this publication were obtained and made available by the MIT-UH-IRTF Joint Campaign for NEO Reconnaissance. The IRTF is operated by the University of Hawaii under Cooperative Agreement no. NCC 5-538 with the National Aeronautics and Space Administration, Office of Space Science, Planetary Astronomy Program. This project was made possible by the outstanding support from the IRTF staff and telescope operators. This paper includes data gathered with the 6.5 m Magellan telescopes located at Las Campanas Observatory, Chile. MIT researchers performing this work were supported by NASA grant 09-NEOO009-0001, and by the National Science Foundation under Grants nos. 0506716 and 0907766. D. Polishook received postdoctoral support at MIT under an AXA Research Fellowship. M. Granvik acknowledges funding from the Academy of Finland (grant 299543). Taxonomic type results presented in this work were determined, in whole or in part, using a Bus-DeMeo Taxonomy Classification Web tool by Stephen M. Slivan, developed at MIT with the support of National Science Foundation Grant 0506716 and NASA Grant NAG5-12355. Over the many years of this project we have benefited from insightful discussions from many colleagues who have further motivated and focused our observational program and science objectives. These individuals include Bill Bottke, Andy Cheng, Beth Clark, Marco Delbo, Alan W. Harris, Alan W. Harris; in that order, Alison Klesman, Amy Mainzer, Joe Masiero, Lucy McFadden, Jessica Sunshine, David Nesvorný, and many more.

A number of observers contributed data to this project, in some cases prior to their publication. We acknowledge them as co-authors in some cases, but also with every attempt at appropriate reference within the full data set included as Appendix I (in print here) and Appendix II (electronic format supplement). The first author accepts full responsibility for any and all omissions or errors. Many individuals participated in this project while students and we acknowledge them here, including (but not limited to): Kimberly A. Barker, Rachel Bowens-Rubin, Tom Endicott, Michael Kotson, Matthew Lockhart, Lindsey Malcom, Megan Mansfield, Mary Masterman, Dimitra Pefkou, Alessandra Springmann, Cavaille Stepanova, Shaye P. Storm. We thank MIT postdoc Alissa M. Earle for assistance in the final revision. We particularly welcome Edward Nelson Binzel and Renata Ava (DeMeo) Morales as new arrivals motivating the completion of this manuscript.

Supplementary material associated with this article can be found, in the online version, at doi:10.1016/j.icarus.2018.12.035.

**Appendix I** Postage Stamp Figure Presentation of Spectral Data.

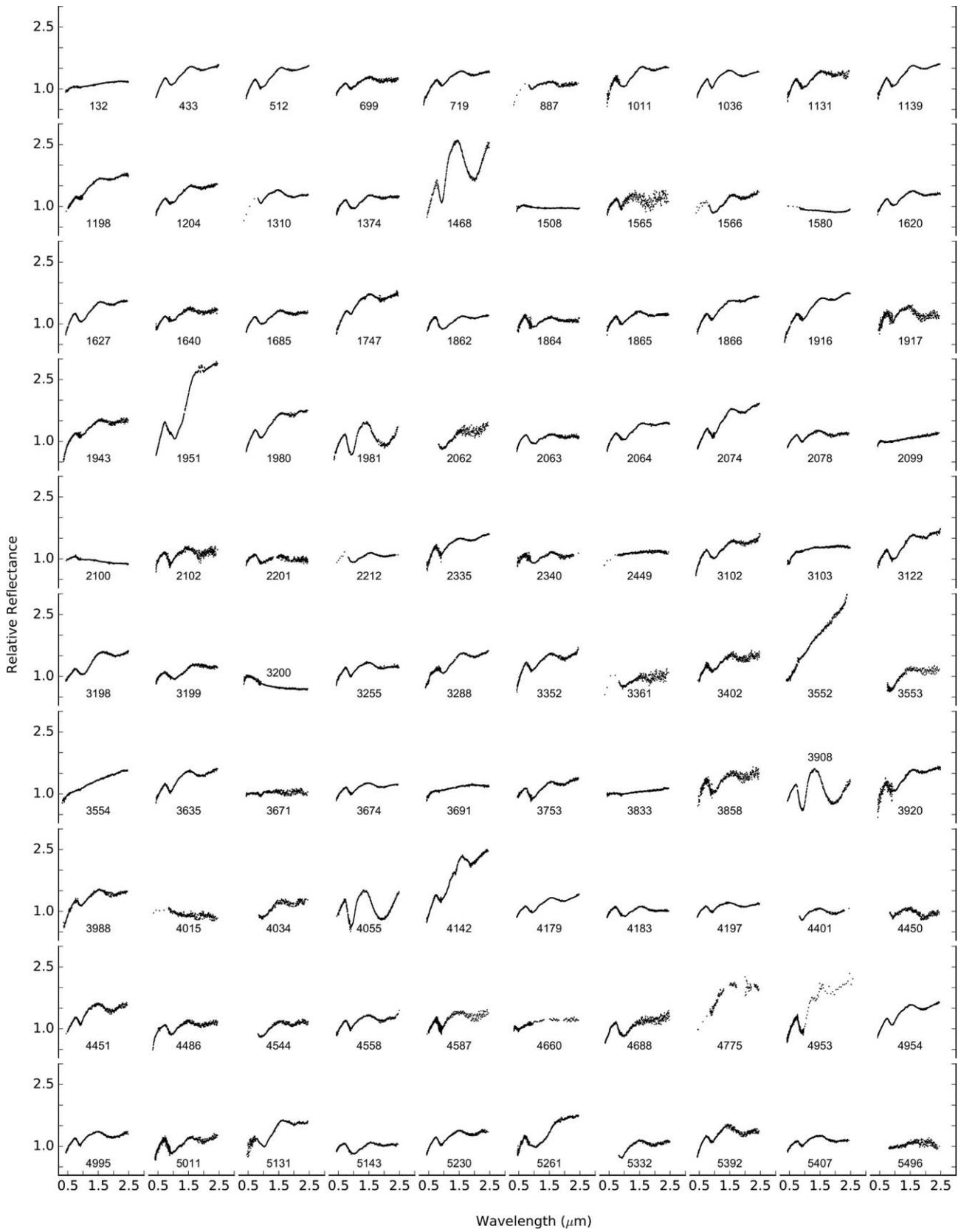

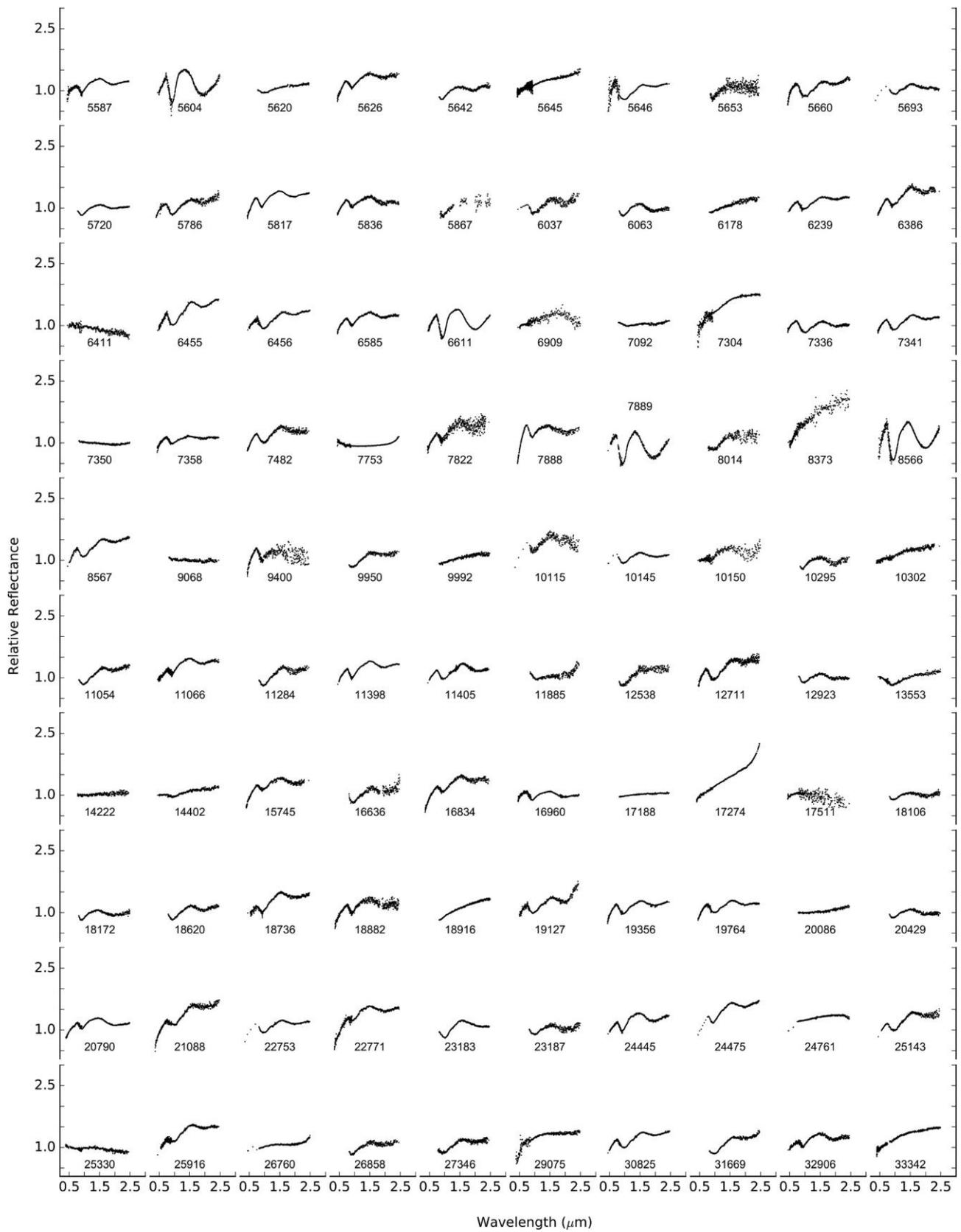

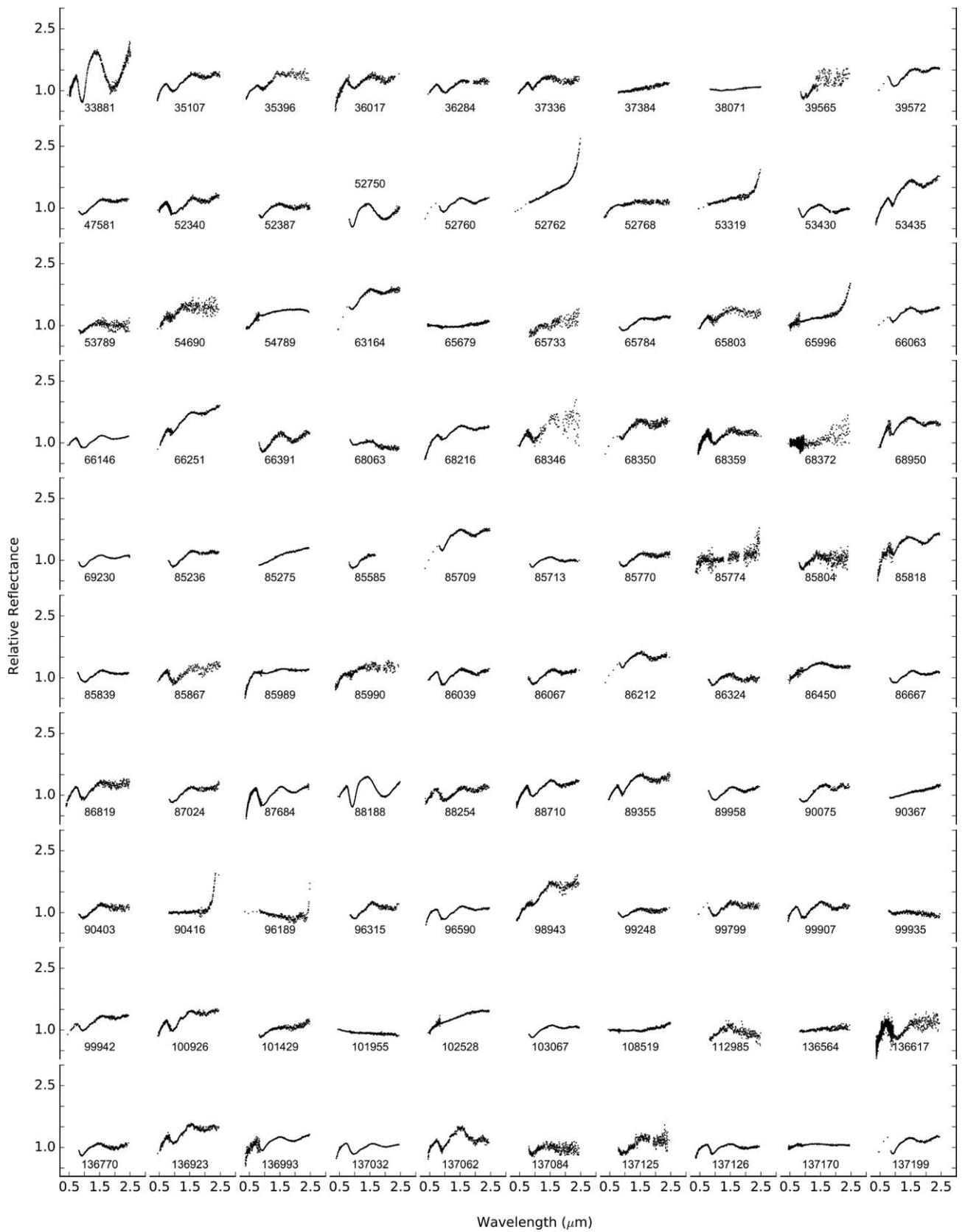

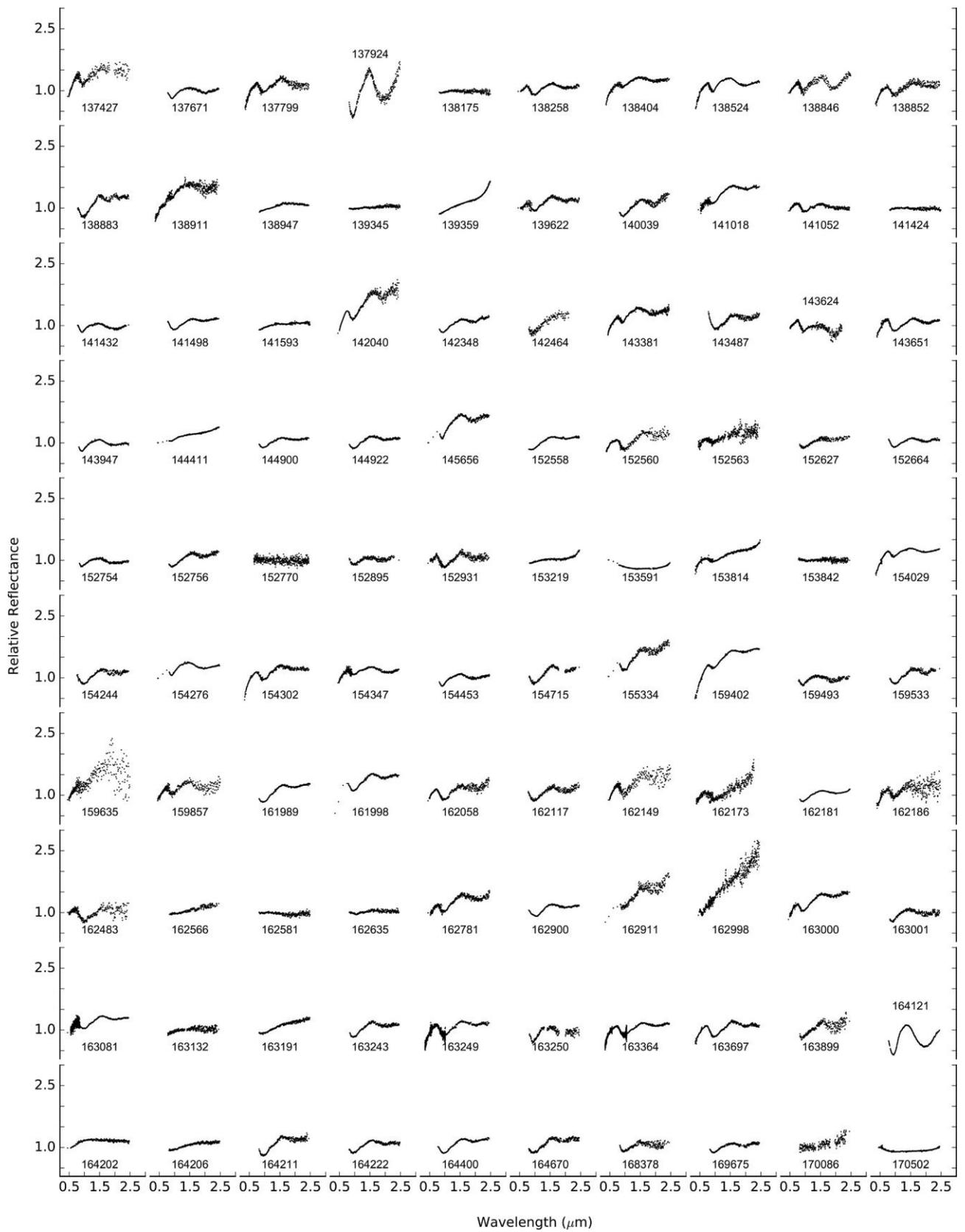

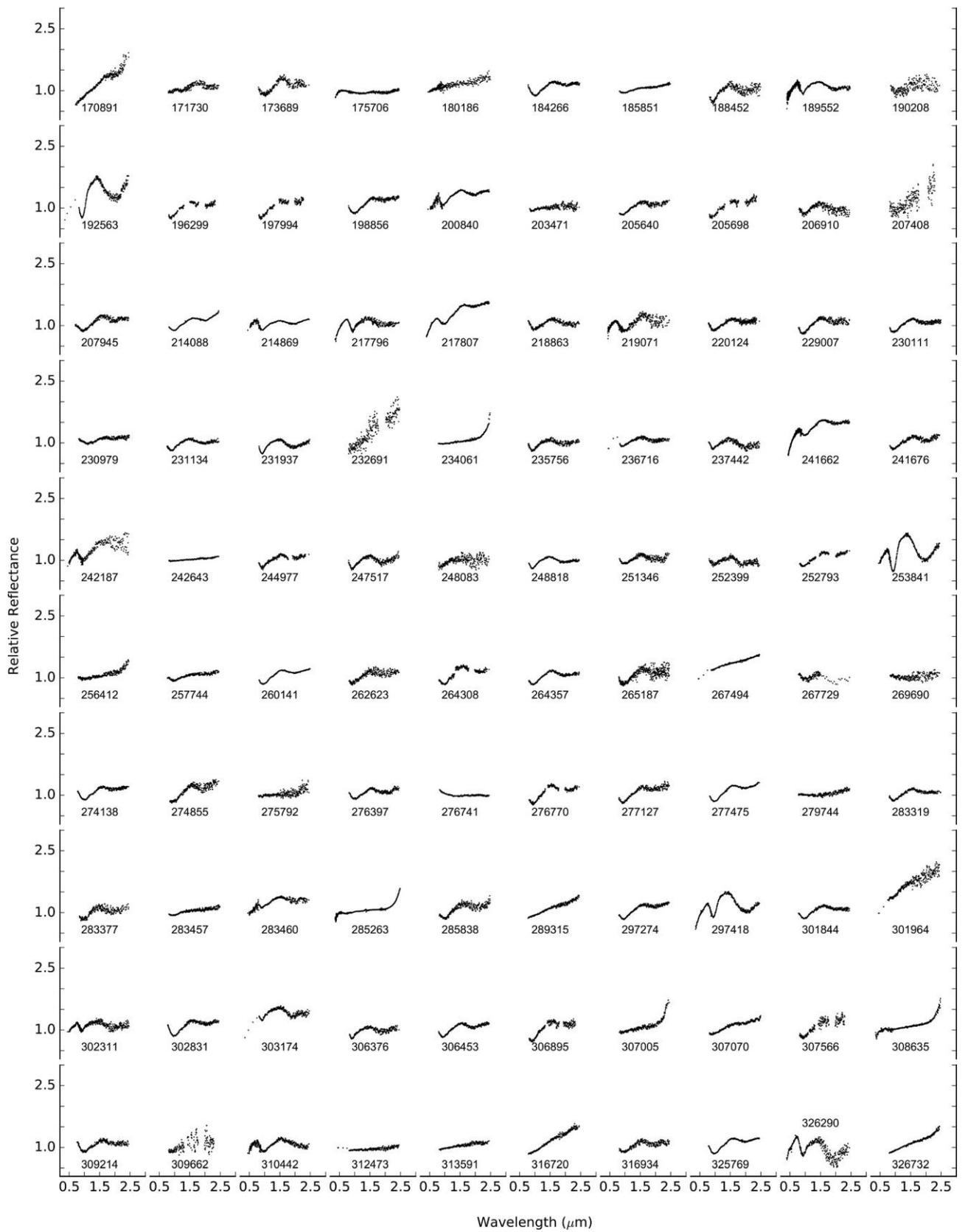

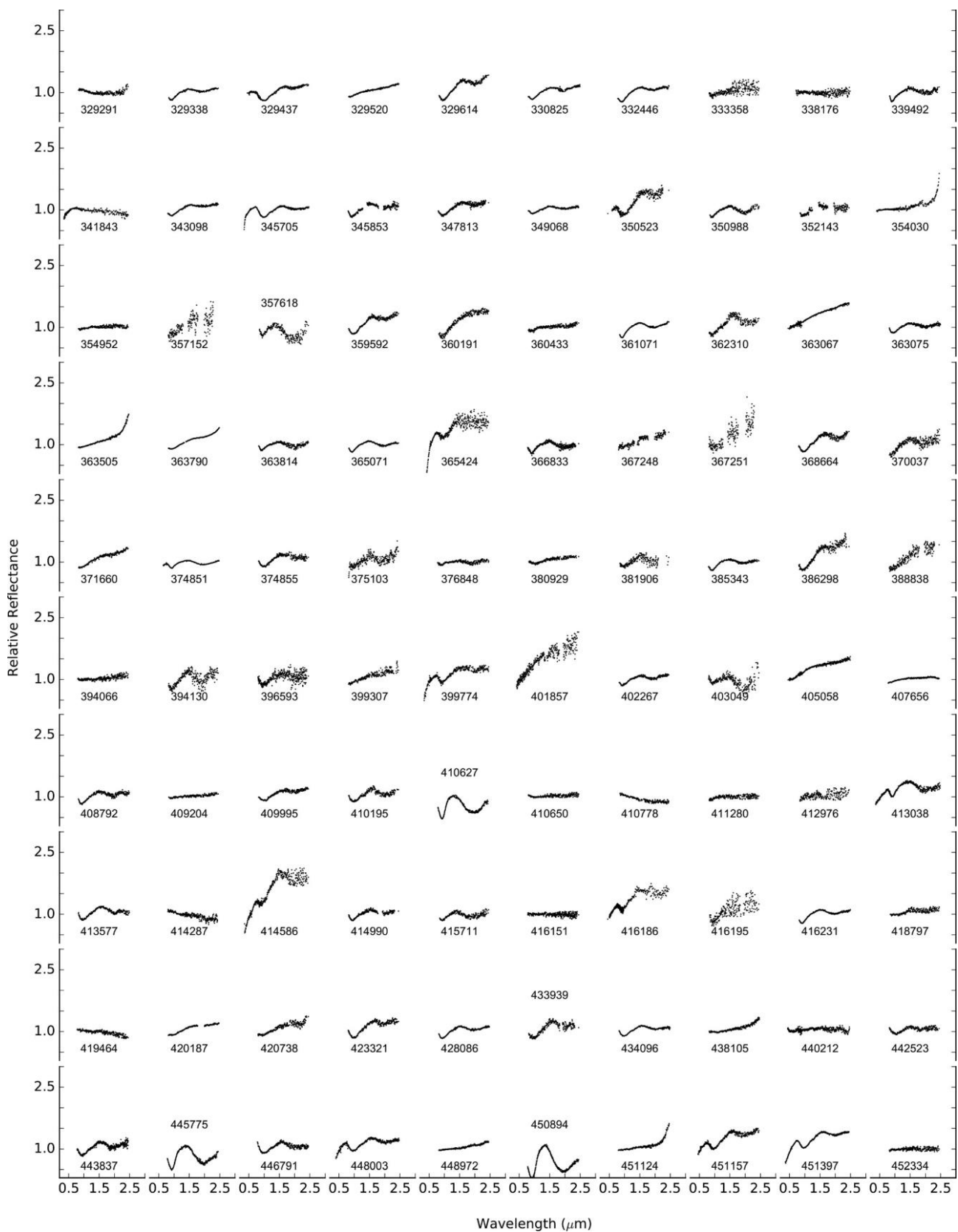

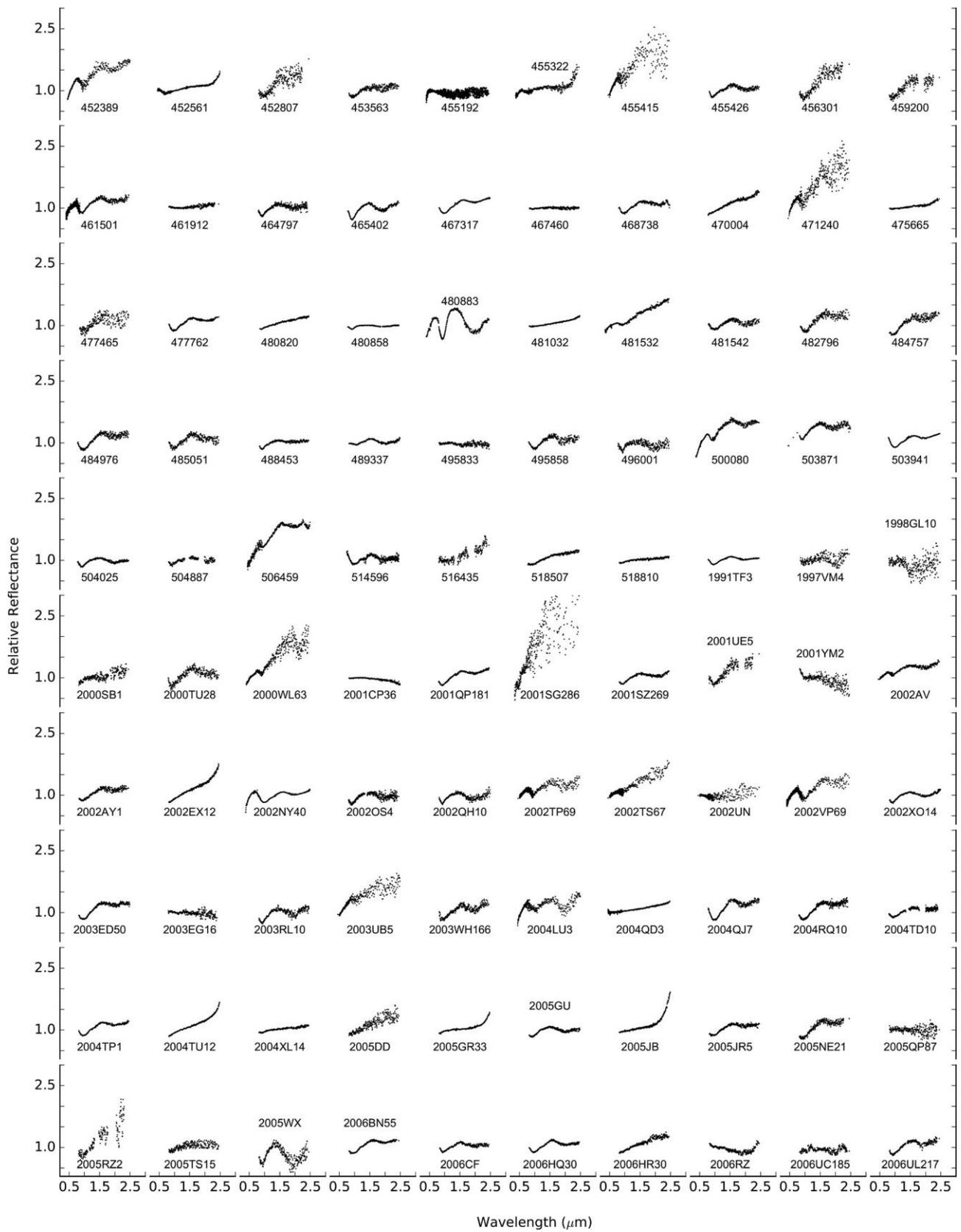

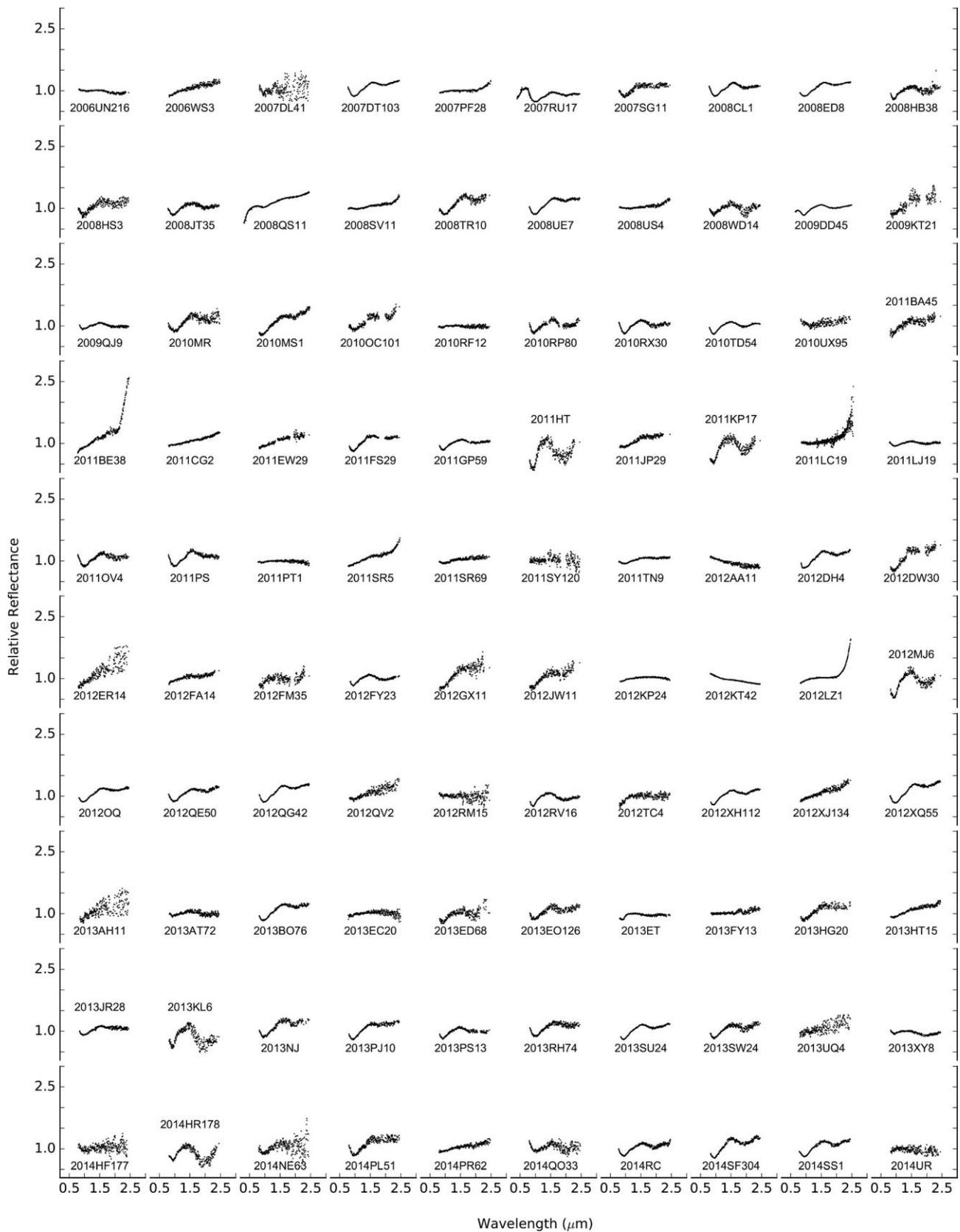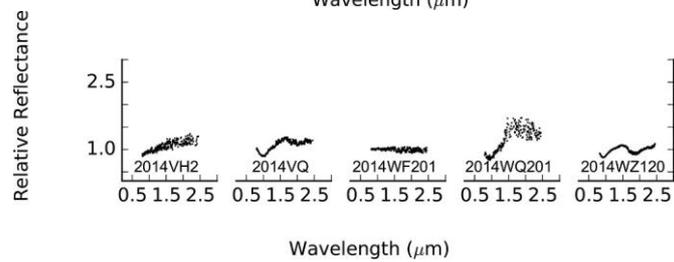